\documentclass[11pt,a4paper]{article}

\usepackage[left=1.0in, right=1.0in, top=1.1in, bottom=1.1in]{geometry}

\usepackage[T1]{fontenc}
\usepackage[latin9]{inputenc}
\usepackage{float}
\usepackage{ifthen}
\usepackage{amsbsy}
\usepackage{amssymb}
\usepackage{amsmath}
\usepackage{amsthm}
\usepackage[]{graphicx}  
\usepackage{setspace}
\usepackage{esint}
\usepackage{comment}
\usepackage{mathtools}
\usepackage{xcolor}
\DeclareMathOperator*{\argmin}{arg\,min} % * Places subscript directly under operator
\DeclareMathOperator*{\argmax}{arg\,max}

\def\ddefloop#1{\ifx\ddefloop#1\else\ddef{#1}\expandafter\ddefloop\fi}
\def\ddef#1{\expandafter\def\csname bb#1\endcsname{\ensuremath{\mathbb{#1}}}}
\ddefloop ABCDEFGHIJKLMNOPQRSTUVWXYZ\ddefloop
\def\ddefloop#1{\ifx\ddefloop#1\else\ddef{#1}\expandafter\ddefloop\fi}
\def\ddef#1{\expandafter\def\csname b#1\endcsname{\ensuremath{\mathbf{#1}}}}
\ddefloop ABCDEFGHIJKLMNOPQRSTUVWXYZ\ddefloop
\def\ddef#1{\expandafter\def\csname c#1\endcsname{\ensuremath{\mathcal{#1}}}}
\ddefloop ABCDEFGHIJKLMNOPQRSTUVWXYZ\ddefloop
\def\ddef#1{\expandafter\def\csname h#1\endcsname{\ensuremath{\widehat{#1}}}}
\ddefloop ABCDEFGHIJKLMNOPQRSTUVWXYZ\ddefloop
\def\ddef#1{\expandafter\def\csname hc#1\endcsname{\ensuremath{\widehat{\mathcal{#1}}}}}
\ddefloop ABCDEFGHIJKLMNOPQRSTUVWXYZ\ddefloop
\def\ddef#1{\expandafter\def\csname t#1\endcsname{\ensuremath{\widetilde{#1}}}}
\ddefloop ABCDEFGHIJKLMNOPQRSTUVWXYZ\ddefloop
\def\ddef#1{\expandafter\def\csname tc#1\endcsname{\ensuremath{\widetilde{\mathcal{#1}}}}}
\ddefloop ABCDEFGHIJKLMNOPQRSTUVWXYZ\ddefloop

\usepackage{prettyref}

\usepackage{nicefrac}
\usepackage{subcaption}
\usepackage{booktabs}

\usepackage{enumitem}

\usepackage{url}
\usepackage[]{hyperref}
\usepackage{cleveref}

\crefname{equation}{}{}
\crefname{proposition}{Proposition}{Propositions}
\crefname{appendix}{Appendix}{Appendices}
\usepackage{autonum}

\usepackage{placeins}
\usepackage{verbatim}

\usepackage[]{color-edits}
\addauthor{vs}{red}
\addauthor{gn}{blue}

\newcommand{\kibitz}[2]{\ifnum\Comments=1{\color{#1}{#2}}\fi}

\newcommand{\R}{\mathbb{R}}

\newcommand{\ba}{\begin{array}}
\newcommand{\ea}{\end{array}}
\newcommand{\bs}{\begin{align}\begin{split}\nonumber}
\newcommand{\bsnumber}{\begin{align}\begin{split}}
\newcommand{\es}{\end{split}\end{align}}

\def\balign#1\ealign{\begin{align}#1\end{align}}
\def\balignat#1\ealign{\begin{alignat}#1\end{alignat}}
\def\bitemize#1\eitemize{\begin{itemize}#1\end{itemize}}
\def\benumerate#1\eenumerate{\begin{enumerate}#1\end{enumerate}}
\newenvironment{talign}
 {\csname align\endcsname}
 {\endalign}
\def\balignt#1\ealignt{\begin{talign}#1\end{talign}}%
\newcommand{\ogd}{\text{OGD}}
\newcommand{\brreg}{\text{BRReg}}
\newcommand{\ftrl}{\text{FTRL}}
\newcommand{\ogdbehav}{\text{OGDBias }}

\newcommand{\regret}{\text{Regret}}

\title{Bid Prediction in Repeated Auctions with Learning}

\date{}

\author{
			 Gali Noti\thanks{Rachel \& Selim Benin School of Computer Science \& Engineering and Federmann Center for
the Study of Rationality, The Hebrew University of Jerusalem, Israel, and Microsoft Research. Supported by the Adams Fellowship Program of the Israel Academy of Sciences and Humanities. This project has received funding from the European Research Council (ERC) under the European Union's Horizon 2020 research and innovation programme (grant agreement No 740282). Email: galinoti@gmail.com.}%
  \and Vasilis Syrgkanis\thanks{Microsoft Research. Email: vasy@microsoft.com}
	}

\begin{document}
\maketitle

\begin{abstract}
We consider the problem of bid prediction in repeated auctions and evaluate the performance of econometric methods for learning agents using a dataset from a mainstream sponsored search auction marketplace. Sponsored search auctions is a billion dollar industry and the main source of revenue of several tech giants. A critical problem in optimizing such marketplaces is understanding how bidders will react to changes in the auction design. We propose the use of no-regret based econometrics for bid prediction, modeling players as no-regret learners with respect to a utility function, unknown to the analyst. We propose new econometric approaches to simultaneously learn the parameters of a player's utility and her learning rule, and apply these methods in a real-world dataset from the BingAds 
%a major 
sponsored search auction marketplace. We show that the no-regret econometric methods perform comparable to state-of-the-art time-series machine learning methods when there is no co-variate shift, but significantly outperform machine learning methods when there is a co-variate shift between the training and test periods. This portrays the importance of using structural econometric approaches in predicting how players will respond to changes in the market. Moreover, we show that among structural econometric methods, approaches based on no-regret learning outperform more traditional, equilibrium-based, econometric methods that assume that players continuously best-respond to competition. Finally, we demonstrate how the prediction performance of the no-regret learning algorithms can be further improved by considering bidders who optimize a utility function with a visibility bias component.
\end{abstract}

\section{Introduction}

Sponsored search auctions are one of the most prominent revenue sources of modern tech giants and among the most profitable electronic marketplaces. Understanding how to design and optimize mechanisms for online ad auctions has been the focus of a long line of work at the intersection of economics and computation, with several streams of research analyzing the design of approximately optimal simple auctions \cite{edelman2007internet,caragiannis2015bounding,lucier2012revenue}, optimizing reserve prices \cite{Mohri2016}, estimation of returns-on-investment \cite{lewis2015unfavorable} and analyzing structural parameters \cite{athey2010structural,Syrgkanis2015}.

An important step in optimizing sponsored search marketplaces is the ability to understand how bidders will respond to market changes or market shocks. One can take a fully unstructured approach to the bid prediction problem by treating it as a time-series forecasting problem. However, such approaches can potentially overfit to the current market design or market setting and will not be able to extrapolate well, when co-variate shifts arise in the data. 

Economic and econometric theory can potentially be beneficial for performing such extrapolation tasks and improve the ability to predict counterfactual behavior. In this paper we perform an empirical evaluation of this statement. The environment that the bidders are facing can be best thought of as a repeated game-theoretic strategic interaction, where the bids of one player affect the reward of another. Thereby, learning models of repeated strategic interactions are most appropriate. One can potentially think of the task as a multi-agent inverse reinforcement learning problem (IRL) \cite{Russel1998,Ng2000,yu2019multi}. However, the players in a sponsored search auction are facing a very complex auction design and a dynamically changing strategic environment. Therefore classical approaches to multi-agent IRL that typically assume that players form consistent beliefs can be problematic. Similar problems arise if one uses more classical economic approaches that assume that the system is at equilibrium \cite{athey2010structural,paarsch2006introduction}, or variations of equilibrium notions that incorporate bounded rationality \cite{Jiang2016}.

An alternative that has received recent attention in the literature at the intersection of economics and computation 
is modelling players as invoking no-regret learning algorithms \cite{Syrgkanis2015,Nisan2017a,Nisan2017b,Braverman2018,Alaei2019}. However, all prior work focused primarily at uncovering structural parameters of the utility of the bidders, such as the value per click, and did not address the empirical performance of such a behavioral assumption in terms of its predictive power. In fact, all prior work on regret-based econometrics fall short in tackling this prediction problem: in order to predict the bid of a player we don't only need to know the parameters of the utility that she is optimizing (e.g. the value-per-click), but we also need to know the ``no-regret'' learning rule, also known as the ``update rule,'' that the bidder is invoking, i.e., how the bidder uses past observations to update her bid so as to minimize regret with respect to her utility over time.

In this work we address this gap: we propose methods for simultaneously learning structural parameters and update-rule parameters from data, and we test whether models of no-regret learning behavior can have predictive power and outperform baselines, by applying our enhanced regret-based econometric methods on a real world sponsored search auction market. We note that even though this theory is making several strong behavioral assumptions on the bidders, we mostly use these assumptions to impose structure on our estimation approach so as to regularize and extrapolate better. The ultimate judge is how well these methods perform in terms of prediction. Thus, in the spirit of George Box's writings \cite{Box1976}, even if these theories could be wrong, our goal is to test whether they could potentially be useful.

Using a large auction dataset from Microsoft's BingAds sponsored search auction marketplace, we show that regret-based bid prediction methods perform comparable to bid-based time-series machine learning baselines when there is no co-variate shift, but outperform these baselines in a statistically significant manner in the presence of a co-variate shift.
We further show that no-regret learning methods outperform methods that rely on the traditional assumption that bidders best reply to competition at every step. 
Moreover, we show that their performance can be  further improved by considering bidders who optimize a utility function with a visibility bias component instead of the standard quasi-linear utility function. The latter gives strong evidence that sponsored search bidders optimize an objective that has an impression-based value component as opposed to only a value-per-click component. This hints at potentially better auction designs that incorporate value-per-impression considerations.

%There have also been recent works on alternative notions of rationality for approaching the bid prediction problem \cite{xu2013predicting,Jiang2016}. In this work we primarily focus on comparing the no-regret model of behavior as compared to classical econometric methods and unstructured time-series machine learning baselines in terms of their predictive power. %and their usefulness when there is a change in the data. 

\section{Bid Prediction in Repeated Ad Auctions} \label{sec:econ}

We consider a repeated sponsored search auction setting. At each period $t$ a set of $n_t$ bidders participate in an auction for advertisement slots that will appear alongside the search results triggered by user queries. We assume that there are $n_t$ bidders and $m$ slots and each slot corresponds to a different probability of receiving a click. Each player $i$ submits a bid $b_t^i$ at each time $t$ and based on the vector of bids, the auction decides an allocation of the slots and a price that each bidder needs to pay. For the goal of our analysis, the actual mechanics of the auction are not important and moreover are too complex to describe, even if proprietary constraints where not at play.

What is important are the following fundamentals. At every period $t$,  %(which in our data analysis will correspond to a period of an hour),
we associate two functions that are sufficient statistics for the strategic reasoning of player $i$: a probability of click curve $x_t^i: \R_+\to [0, 1]$ and a cost-per-click (CPC) curve $p_t^i: \R_+\to \R_+$, which for every bid $b$ return the average probability of receiving a click and the average cost-per-click for the auctions that occurred during that period had the player submitted a bid $b$. In practice, such curves are reported to the bidders at periodic intervals through revenue optimization feedback tools provided by sponsored search auction marketplaces. Given that from now on we will be mostly focusing on the perspective of a single bidder, we will drop the index $i$. The competition stemming from other bidders is summarized in the sufficient statistic of the cost and click curves. 
Our main question is the following:

\emph{Given the history of play up till time $t$, can we forecast for an advertiser the future time series $b_{t+1:T}$? Moreover, can we forecast this series when there is a change in the market at time $t$ (a co-variate shift)?}

% \emph{Given a %time series 
% sequence of bids $b_{1:t}$ for an advertiser,  can we forecast the future time series $b_{t+1:T}$? Moreover, can we forecast this series when there is a change in the market at time $t$ (a co-variate shift)?}

%% Version 1: 
%For each of these variants of the question we will consider two forecasting tasks. 
%In the first, we are asked to produce a future time-series $b_{t+1:T}$, %(series prediction), 
%where the prediction of bid $b_{\tau}$, for $\tau>t$, is performed solely with knowledge of the bids $b_{1:t}$ and the past cost and click curves $x_{1:t}$, $p_{1:t}$, and the %future 
%cost and click curves up till time $\tau-1$, i.e. $x_{t+1:\tau-1}$, $p_{t+1:\tau-1}$.\footnote{Typically forecasting cost and click curves is an easier task, since they only depend on aggregate market statistics, hence for our structural econometric methods we will assume that such forecasts are available. %It is a fruitful avenue for future research whether we can endogenize the prediction of the future cost and click curves, but such a task was not feasible with the data that were available to us.
%} 
%In the second, we will consider a one-step-ahead prediction, where we forecast $b_{\tau}$ from bids $b_{1:\tau-1}$ and curves $x_{1:\tau-1},p_{1:\tau-1}$.

% Version 2: 
For each of these two variants of the question we will consider two forecasting tasks. 
In the first, we consider a one-step-ahead prediction, where for each future time $\tau>t$ we forecast $b_{\tau}$ 
based on all data till time $\tau$, which include the bids $b_{1:\tau-1}$ and the curves $x_{1:\tau-1}$ and $p_{1:\tau-1}$.
In the second, we are asked to produce a future time-series $b_{t+1:T}$, %(series prediction), 
where the prediction of bid $b_{\tau}$ is performed solely with knowledge of bids up till time $t$, i.e., $b_{1:t}$, and the past cost and click curves up till time $\tau-1$, i.e., $x_{1:\tau-1}$, $p_{1:\tau-1}$. The second series-prediction task is the more difficult task since error may be accumulated with time. %is similar to training an agent based on all data up to $t$, and then ``releasing'' the agent to bid against reality. 

\section{No-Regret Learning and Structural Econometrics} \label{sec:interpolating_methods}

We will consider a structural econometric approach to the bid prediction task, invoking techniques from classical auction theory and econometrics in auctions (see, e.g., \cite{paarsch2006introduction}). %Since 
While most classical econometric theory in auctions tackles static auction settings or imposes very strong dynamic equilibrium conditions, we will primarily focus on the recent line of work at the intersection of economics and computer science, that models players as no-regret learners and performs econometrics under such a behavioral assumption \cite{Nekipelov2015,Nisan2017a,Nisan2017b}. In this section we describe the assumptions that these structural methods are making and how to transfer these assumptions to an estimation and prediction strategy.

In order to understand how bidders will behave in the future in a model-based manner, one needs to first 
model the objective that the bidders are optimizing and second 
the approach that they use to optimize over time and handle uncertainty. %(how other bidders will behave, which new bidders will arrive, etc). 
A standard assumption in auction theory is that players have a utility from each auction that takes the form: 
$u_t(b; v) = (v - p_t(b))x_t(b)$, 
%\begin{equation}
%    u_t(b; v) = (v - p_t(b))\, %x_t(b) 
%\end{equation}
i.e., the utility is the expected number of clicks times the value-per-click (VPC) $v$ minus the expected payment. We will adopt such a quasi-linear utility function. 
Thus, the only parameter that we need to estimate from the data for each player is the value-per-click $v$.

A classical framework in machine learning on repeated decision making in the face of uncertainty is that of no-regret learning. The no-regret learning framework posits that bidders will choose a bid $b_t$ at each period, such that their regret against submitting any fixed bid in hindsight vanishes to zero as they play for more and more periods, i.e.:
\begin{equation}
    \regret(u_{1:T}, b_{1:T}; v) = \sup_{b} \frac{1}{T} \sum_{t=1}^T (u_t(b; v) - u_t(b_t; v)) = o_p(1)
\end{equation}
We will adopt the regret framework and assume that bidders use some form of no-regret algorithm to optimize their bid. %We note that 
Contrary to the no-regret framework, traditional econometrics in auctions typically assumes that players best respond to the competition, i.e., $b_{t+1}=\argmax_{b} u_t(b; v)$, and use this property to identify the value $v$ (see, e.g., \cite{athey2010structural} for such an econometric approach applied to  sponsored search auction data). However, we will see in the empirical part that such a best-response (BR) algorithm is outperformed by no-regret based algorithms in terms of their predictive power.

Since we are not only interested in uncovering structural parameters of the setting, but also in predicting future behavior, we will consider several classes of no-regret algorithms and %fit their parameters to data.
will show how to learn their update-rule parameters together with the structural parameters from the data. 
We will make the assumption that the utility of the player is a concave function of the bid, which renders the problem that the bidder is facing an online convex optimization problem \cite{Zinkevich2003,shalev2012} in one dimension. Thus, we will consider several widely studied algorithms for this setting. 
In each of these algorithms we will describe the update rule $h(\cdot)$  (the next bid of a player as a function of past bids, click curves and cost curves) and provide some context on where this update rule stems from.

%\begin{enumerate}[topsep=0pt,itemsep=-1ex,partopsep=1ex,parsep=1ex]
 
\textbf{Online Gradient Descent (OGD) \cite{Zinkevich2003}:}
We mainly focus on the OGD algorithm that updates the bid at every step as follows:
    \begin{align}
    b_{t+1} = h_{\ogd, \eta, v}(b_t, u_t) := b_t + \eta \nabla_{b} u_t(b_t; v)
    \end{align}
    OGD can be thought of as regularized best response with momentum, with respect to the last-period ``linearized'' utility $\tilde{u}_t(b; v) = \nabla_{b} u_t(b_t; v) \cdot (b - b_t)$, i.e.: 
    %\begin{align}
    $b_{t+1} = \arg\max_{b} \tilde{u}_t(b; v) - \frac{1}{2\eta} \|b-b_t\|_2^2$. 
    %\end{align}
    Moreover, it can also be thought of as regularized best response to the past average of linearized utilities, with shrinkage bias:
    %\begin{align}
        $b_{t+1} = \arg\max_{b} \sum_{\tau=1}^t \tilde{u}_\tau(b; v)  - \frac{1}{2\eta} \|b\|_2^2$.
    %\end{align}
		Thus, two key elements that distinguish between OGD and the best-reply (BR) algorithm are the consideration of the full history rather than the previous step only, and the addition of a regularization term.
    
 \textbf{Implicit OGD (BR-Reg):} $b_{t+1}$ is defined as the solution to the equation:
    %\begin{equation}
    $b_{t+1} - b_t = \eta \nabla_{b} u_t(b_{t+1}; v)$. 
    %\end{equation}
    This is also referred to as the \emph{implicit gradient descent} \cite{toulis2017}. This algorithm has the same interpretation as OGD of a regularized best-response with momentum, but without the linearization of the utility component, i.e., it is equivalent to:
    \begin{align}
        b_{t+1} = h_{\brreg, \eta, v}(b_t, u_t) := \arg\max_{b} u_t(b; v) - \frac{1}{2\eta} \|b - b_t\|_2^2
    \end{align}
   
 %\textbf{FTRL without linearization and with recency bias:}
 \textbf{Follow the Regularized Leader (FTRL):} FTRL without linearization and with recency bias updates the bid at every step as follows:
    \begin{equation}
        b_{t+1} = h_{\ftrl, \eta, v}(u_{1:t}) := \arg\max_{b} \sum_{\tau=1}^t \beta^{t-\tau} u_\tau(b; v)  - \frac{1}{2\eta} \|b\|_2^2
    \end{equation}
    This implies that $b_{t+1}$ is defined as the solution to the equation:
    %\begin{equation}
      $  b_{t+1} = \eta \sum_{\tau=1}^t \beta^{t-\tau} \nabla_{b} u_\tau(b_{t+1}; v)$.
    %\end{equation}
    Recency bias has been analyzed in the context of no-regret algorithms \cite{Fudenberg2014} and relates to learning in changing environments \cite{Hazan2009, Adamskiy2012} and fast convergence in games \cite{Syrgkanis2015}. In the FTRL implementation we use $\beta=0.9$. We will also refer to the special case where $\beta=1$ and $\eta=\infty$ as the  \textbf{Follow-the-Leader algorithm (FTL)}, which is also a no-regret algorithm when the utility functions are strongly concave (see, e.g., \cite{Hazan2007strongly}).\footnote{It is possible to view the counterfactual-curve based algorithms, in which players respond to their estimates on the other players' average behavior, as mean-field algorithms \cite{meanfield2014} with no additional constraints. The most basic example would be the BR algorithm which is a simple best response to the last observed click and cost curves. We have also evaluated FTL with recency bias, i.e., the special case of FTRL with $\eta=\infty$ and $\beta=0.9$ that can be viewed as mean field where the beliefs of the players are estimated by the running averages of recent curves. However, in the paper we present the basic variant of FTL with $\beta=1$ (i.e., with no recency bias) that achieves better results on our empirical dataset. }

\begin{figure*}[t]
\centering
\begin{subfigure}{.49\linewidth}
\includegraphics[width=1.00\linewidth]{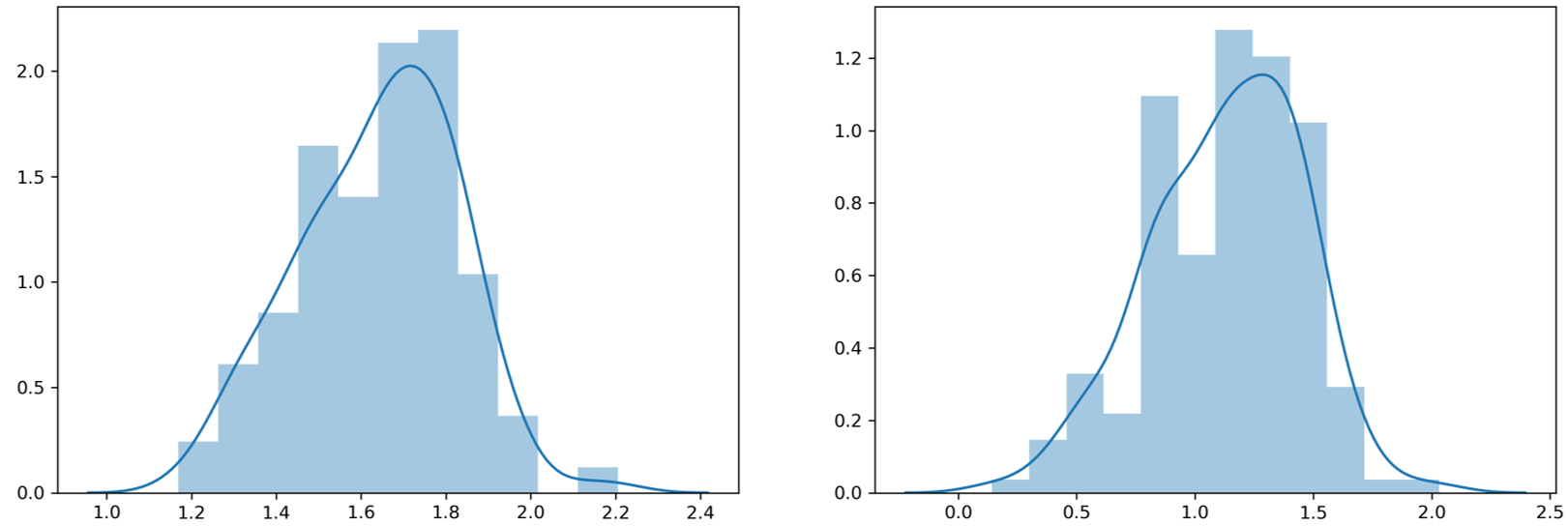}
\caption{Estimated value over mean bid (bid shade ratio).}
\label{fig:shade_ratio}
\end{subfigure}
\begin{subfigure}{.49\linewidth}
\includegraphics[width=1.00\linewidth]{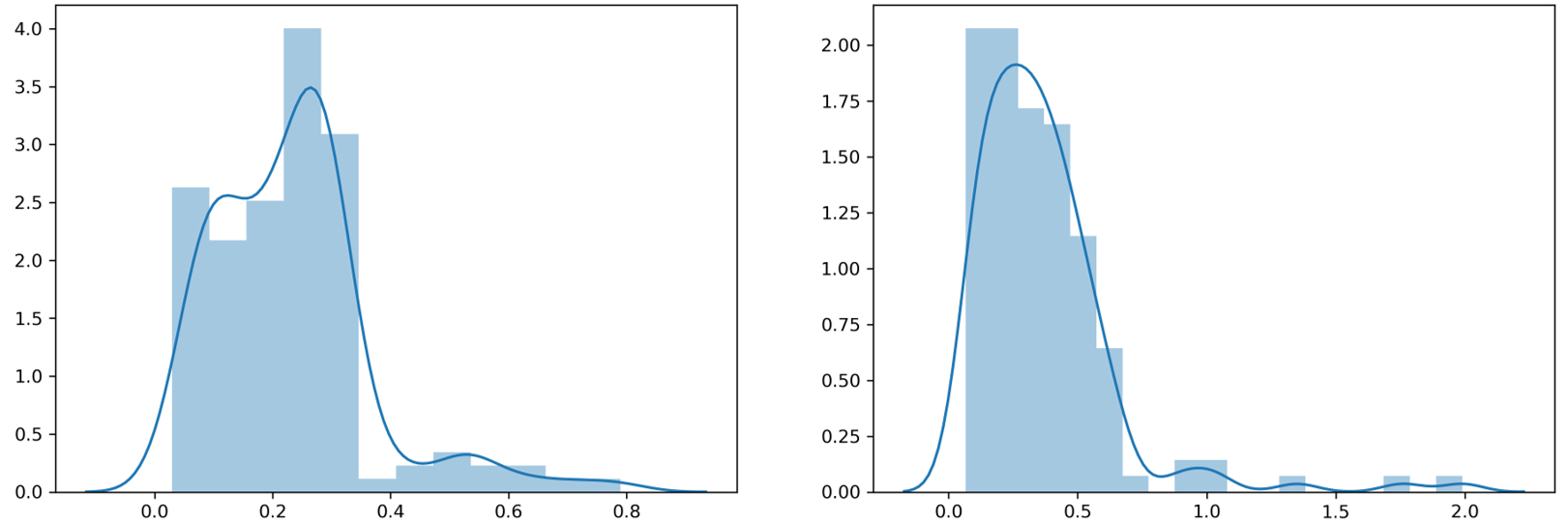}
\caption{Coefficient of variation of daily estimated values.}
\label{fig:daily_values_dist}
\end{subfigure}
\caption{Comparison of min regret (a \& b right) vs. quantal regret (a \& b left) value estimates.}
\label{fig:min_v_quantal}
\end{figure*}

\paragraph{Algorithm-independent VPC estimation:} We will estimate the VPC $v$ of the player solely based on the no-regret condition and irrespective of the update algorithm. %We found that this leads to much more stable and intuitive value estimates. 
We consider the value estimation algorithms proposed in \cite{Nekipelov2015,Nisan2017b}. The basic approach one could take (as described in \cite{Nekipelov2015}) is to choose the parameter $v$ that achieves the smallest possible regret level for the player, referred to as the \emph{min-regret estimate}, i.e.:
%\begin{equation}
    $v_{mr} = \argmin_{v \in V} \regret(u_{1:T}, b_{1:T}; v)$. 
%\end{equation}
A more stable alternative was provided in \cite{Nisan2017b} that propose the use of a soft-min version of min regret, referred to as the \emph{quantal-regret estimate}:
\begin{equation}
    v_{qr} = \frac{\sum_{v \in V} v \exp\{-\lambda \regret(u_{1:T}, b_{1:T}; v)\}}{\sum_{v \in V} \exp\{-\lambda \regret(u_{1:T}, b_{1:T}; v)\}}
\end{equation}
where $V$ is a set of candidate valuations.\footnote{In practice, for each bidder we take $V$ to be a grid of values ranging from $1\%$ of the bidder's average bid to $6$ times her average bid.} The authors also provide a Bayesian justification of this choice as imposing a prior on the space of valuations.

We found empirically that the quantal-regret value estimate is more stable and less sensitive to estimation errors than the min-regret estimate, and thus we use the quantal-regret value estimate in our empirical analyses. Figure~\ref{fig:min_v_quantal} justifies the use of $v_{qr}$ over $v_{mr}$ on our data, as it leads to more reasonable predictions on how much players shade their bid (i.e., what fraction of their value is their bid), and how their valuation varies across days of the week, if we were to learn a separate value on solely the dataset of each day. Moreover, we find that the bid difference is positively correlated with the recent gradient of the utility evaluated at the quantal-regret estimate, as predicted by the OGD algorithm, providing further justification for the use of the quantal-regret value (see %Figure ~\ref{fig:grad_v_diff_ogd} 
Appendix \ref{app:qr_values}).

\paragraph{Algorithm-specific step-size estimation:} The OGD, BR-Reg and FTRL algorithms contain a step-size parameter $\eta$ that intuitively controls how aggressively the algorithm responds to new evidence. %i.e. how much does the bid change after every period. 
We will estimate %the step-size 
$\hat{\eta}$ of each algorithm from the data, given our estimated VPC, by minimizing the mean squared prediction error on the training set: 
%$\argmin_{\eta} \frac{1}{t}\sum_{\tau=1}^t (b_{\tau} - h(b_{1:\tau-1}, u_{1:\tau-1}))^2$, 
$\argmin_{\eta} \frac{1}{t-1}\sum_{\tau=1}^{t-1} (b_{\tau+1} - h(b_{1:\tau}, u_{1:\tau}))^2$, 
where $h$ is the algorithm update rule. Since this is a scalar parameter, in the worst-case the optimization requires a grid search. Observe that for OGD, finding $\eta$ that minimizes the mean squared prediction error on the training set:
\begin{equation}
    \hat{\eta}_{\ogd} := \argmin_{\eta \in \R_+} \frac{1}{t-1}\sum_{\tau=1}^{t-1} \left(b_{\tau+1} - h_{\ogd, \eta, v_{qr}}(b_\tau, u_\tau)\right)^2
%\end{equation}
%\begin{equation}
 := \argmin_{\eta \in \R_+}\frac{1}{t-1}\sum_{\tau=1}^{t-1} \left(b_{\tau+1} - b_\tau - \eta \nabla_{b} u_\tau(b_\tau)\right)^2
\end{equation}
is equivalent to finding the $\eta$ from the linear regression of $b_{\tau+1}-b_\tau$ on $\nabla_b u_\tau(b_\tau)$.\footnote{In practice, we enforce positivity of $\eta$ by returning the absolute value of the unconstrained optimal solution.} For %the other algorithms that we consider 
BR-Reg and FTRL 
we perform a grid search to solve the optimization problems.
% \begin{align}
%     \hat{\eta}_{\brreg} :=~&\argmin_{\eta \in \R_+} \frac{1}{T}\sum_{t=1}^T \left(b_{t+1} - h_{\brreg, \eta, v_{qr}}(b_t, u_t)\right)^2\\
%     \hat{\eta}_{\ftrl} :=~& \argmin_{\eta \in \R_+} \frac{1}{T}\sum_{t=1}^T \left(b_{t+1} - h_{\ftrl, \eta, v_{qr}}(u_{1:t})\right)^2
% \end{align}

\section{Data Description}  \label{sec:data}
% Summary statistics of the data and description.

\begin{figure*}[t]
\centering
\begin{subfigure}{.32\linewidth}
\includegraphics[width=1.00\linewidth]{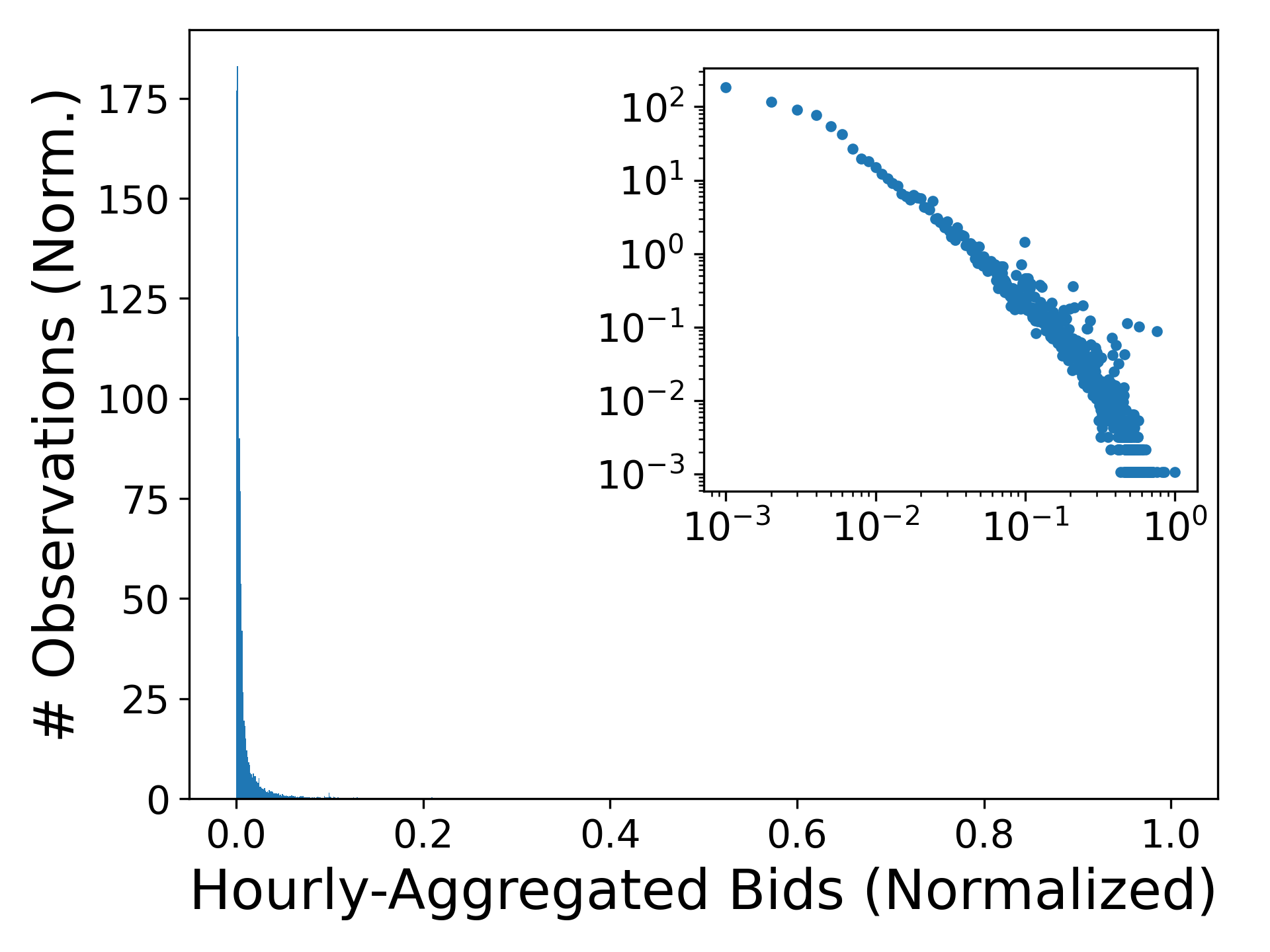}
\caption{}  
\label{fig:bid_dist}
\end{subfigure}
\centering
\begin{subfigure}{.32\linewidth}
\includegraphics[width=1.00\linewidth]{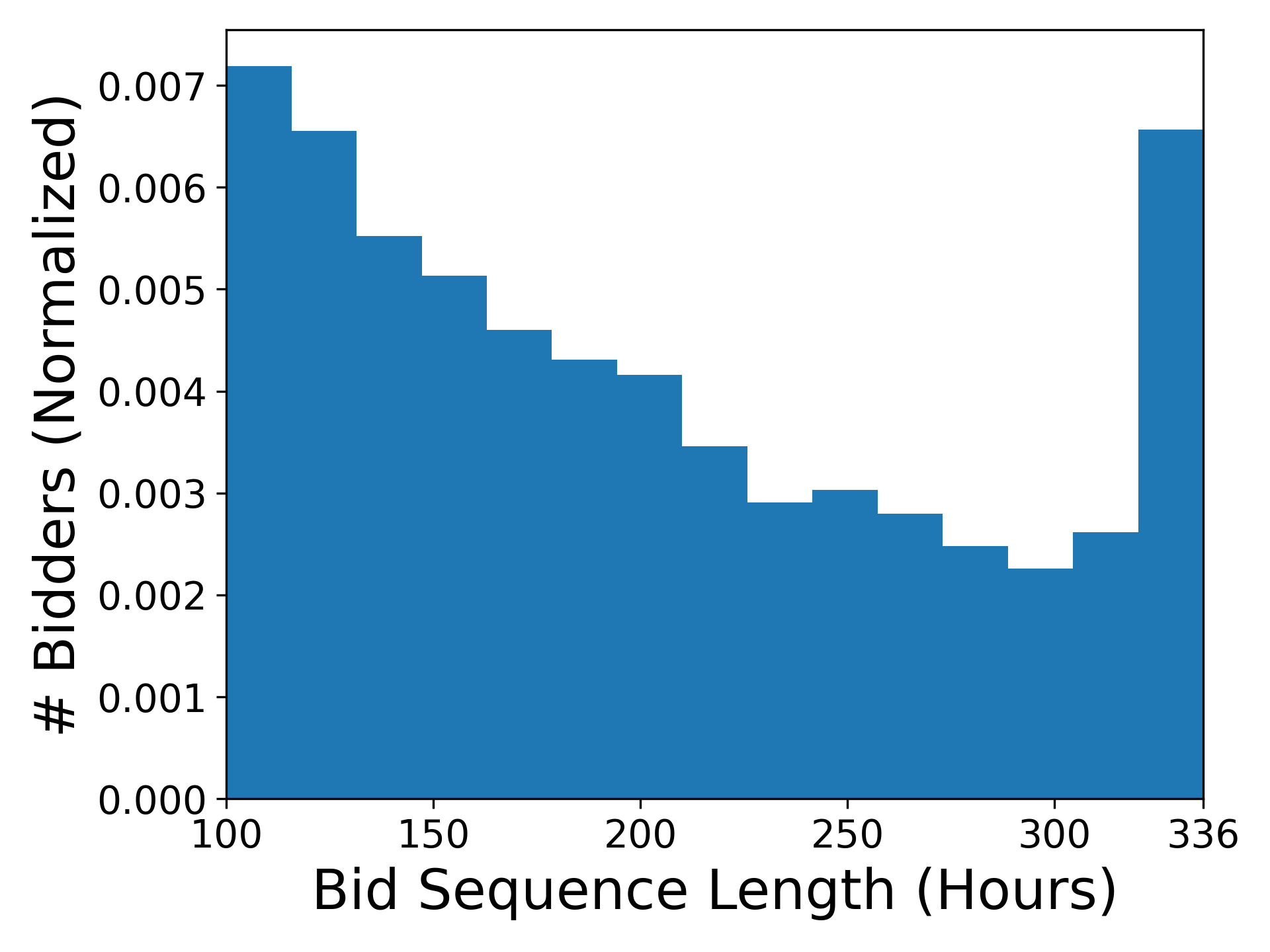}
\caption{}  
\label{fig:sequence_length_dist}
\end{subfigure}
\centering
\begin{subfigure}{.32\linewidth}
\includegraphics[width=1.00\linewidth]{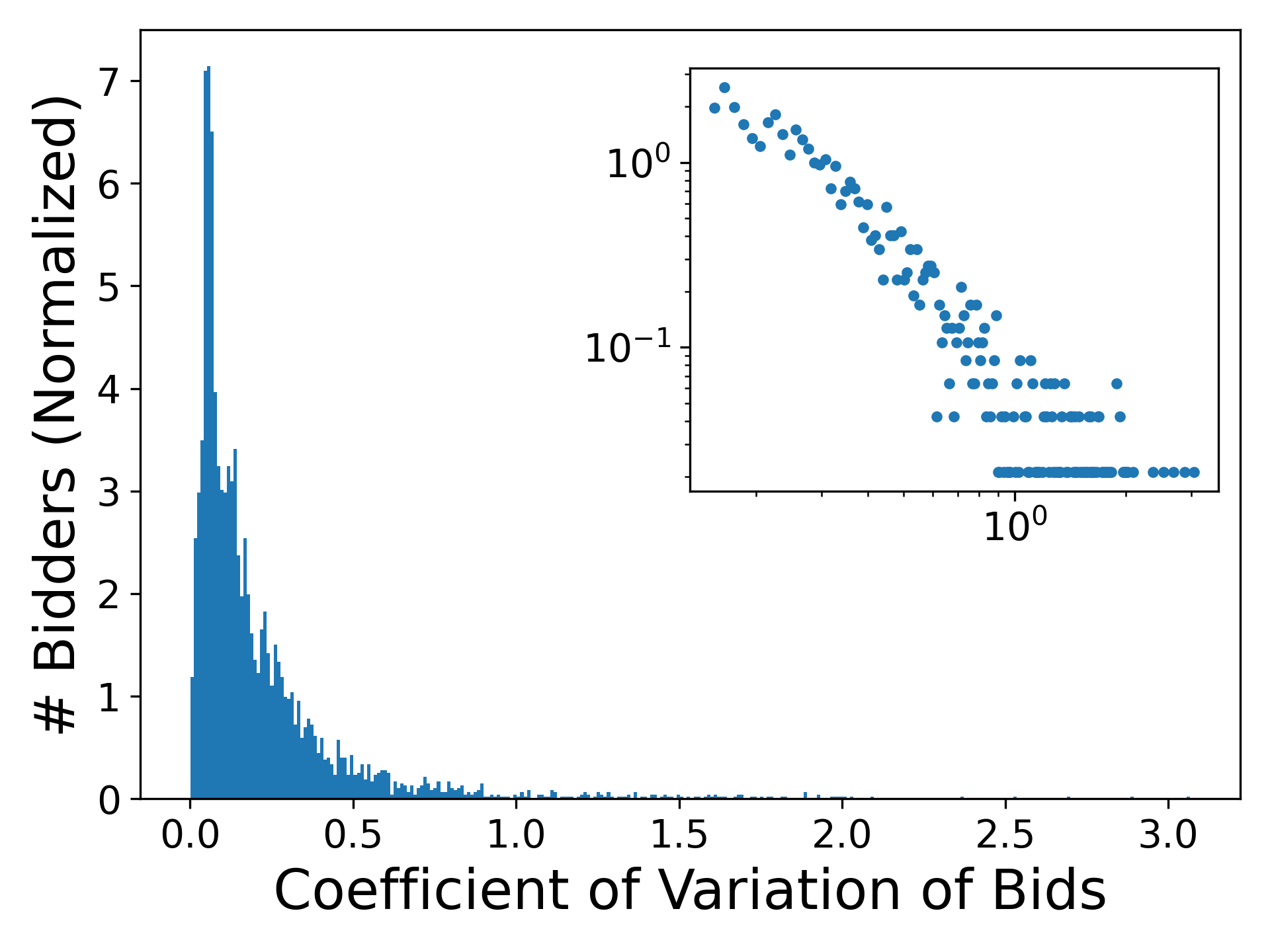}
\caption{}  
\label{fig:insample_coefficient_variation}
\end{subfigure}
\caption{The Dataset. }
\label{fig:insample_setting}
\end{figure*}

  We analyze sponsored-search auction data from the BingAds sponsored search auction marketplace.
  The dataset consists of bidding data for 13 high-volume keywords collected in a period of two weeks. For each bidder, we analyze data that include the bids the player made, the bidder's cost-per-click (CPC), the slot and the clicks that the bidder won in each auction, as well as counterfactual data of CPC and click rates that the bidder could have obtained for different counterfactual bids according to the competition in the auction. These counterfactual information was generated via the Genie system \cite{Genie2019}. For each bidder and auction the bidder participated in, 
  the counterfactual data consist of CPC and probability of click points for counterfactual bids according to 12 different multipliers of the actual bid the bidder made: $(0.1, 0.2, 0.4, 0.6, 0.8, 1.0, 1.2, 1.4, 1.6, 1.8, 2.0, 5.0)$.
    
 Typically, bidders participate in multiple auctions every hour, and we aggregate the data by hour for each bidder. %to obtain a sequence of length 336 hours, possibly with null entries for hours where the player did not participate in any auction. 
The hourly aggregated bid of a player is %simply 
the average of bids the player placed during that hour. 
The hourly counterfactual data for each bidder is a collection of the 12-point discrete curves of each of the auctions the player participated during the hour. We aggregate these hourly curves for each player by fitting the click data points to a concave function of the form $\frac{ax}{b+x}$, which gave the best out-of-sample MSE in a 5-fold cross validation on a validation keyword among the functional forms we tested (including a linear function, a sigmoid $\frac{a}{1 + e^{-b (x - c)}}$ and a convex function $ax^2 + bx^4$).
The CPC data points are fitted to a linear function $ax$ to ensure concave utilities. 
This is of course a modeling assumption that is put to the test in the prediction performance of the economic models.
%and the CPC data points are fitted to a linear function $ax$. 	These functional forms gave the best out-of-sample MSE in a 5-fold cross validation on data of a validation keyword among the functional forms we tested, which in addition include a sigmoid $\frac{a}{1 + e^{-b (x - c)}}$ and a convex function $ax^2 + bx^4$. 
%Fitting the counterfactual data to smooth functions is required to produce continuous and differentiable utility curves for the economic analysis. Of course, this is a modeling assumption that is put to the test in the prediction performance of the economic models.
The dataset that we analyze consists of data of players who participated in auctions in at least $100$ hours and won the top position at least once during the whole two weeks of data. In addition, we require that the players place only positive bids and have non-zero variance in both their train and test data 
(see details on data splitting in Section \ref{subsec:insample_dataset}).

The full dataset, after the filtering described above, includes data of multiple thousands of bidders\footnote{Throughout the paper, we omit the exact numbers %sensitive information 
due to proprietary constraints. 
We also note that the data we analyze went through a strict anonymization process before being provided for this research to protect user privacy, and it does not include any personal or identifiable information.}, %with a total of $934,324$ active hour data and an average of
with an average of 
$202.4$ active hours per bidder, which aggregate in total a number of  auctions of the order of multiple millions. 
Figure \ref{fig:bid_dist} shows the normalized distribution of  hourly-aggregated bids across all players in the dataset. The bidding levels are diverse and span several orders of magnitudes, with, roughly, a power-law  distribution, as seen in the logarithmic scale inset. %The diversity of bids is one of the factors that makes learning a model per-player more effective than the possibility of learning between players. 
Figure \ref{fig:sequence_length_dist} shows the distribution of sequence lengths across bidders.

Figure \ref{fig:insample_coefficient_variation} shows the distribution of coefficient of variation of bids across bidders. 
%The coefficient of variation quantifies the variability of the bids a player made relative to their average magnitude. 
The most frequent bidding behavior is of moderate variability of around $5$\% deviation from the average. However, as seen in the median at $12.8$\%, the majority of players have high variability bids. The average is as high as $21.2$\%, due to a non-negligible number of bidders with extremely high variability, as seen by the power-law tail distribution; see the inset in logarithmic scale of players with coefficient of variation of above $0.15$.
	%peak of the distribution is at $0.047$, the median is $0.128$, and the average is $0.212$ (higher than the median due to non-negligible number of bidders with high bid variability, as seen in the distribution tail). 

\section{Predictive Performance in the In-Sample Setting} \label{sec:insample}

In this section we evaluate the predictive performance of the econometric-based methods in a setting in which the train and test data have similar distributions, which we refer to as the  {\em in-sample} setting.  We show that in this setting, unstructured bid-based machine-learning (ML) benchmarks do well and that the 
structured econ OGD method manages to achieve comparable results.
%to the ML methods that indeed do well in this setting. 
We also observe large differences between the best-reply (BR) and the no-regret OGD methods. In Section \ref{subsec:insample_econ} we evaluate the performance of the econ methods discussed in Section \ref{sec:interpolating_methods}. %that interpolate between BR and OGD.
The results show that the no-regret methods outperform those that are not regret minimizing.

\subsection{The In-Sample Prediction Setting} \label{subsec:insample_dataset}

We analyze the dataset described in Section \ref{sec:data}. 
We use data of one of the 13 keywords for development and validation and exclude this keyword from the prediction analysis. The dataset without the validation keyword consists of $96.2$\% of the players in the full dataset.
For each player we divide the bid sequence and use the first 90\% as training data and the last 10\% for test. Training sequence lengths range between $90$ and $303$ hours with an average of % $182.6$ (with aarp)
$182.2$ hours,    % without aarp
and test sequence lengths range between 10 and 33 hours, with an average of %$19.8$ (with aarp)
$19.7$ hours. 
As explained in Section \ref{sec:econ}, we evaluate the predictive performance of the methods both in {\em series} prediction, where each model is trained on the training sequence of a player and then remains fixed for the prediction phase on the entire test sequence, and in a {\em stepahead} prediction task, where the models are re-fitted on the true data at every step and predict only a single step at a time. 

%\subsection{Benchmark Machine-Learning (ML) Methods} \label{subsec:ml_methods}
\subsection{Machine-Learning (ML) Benchmarks} \label{subsec:ml_methods}

We implemented the following ML benchmarks: A linear model %with a combination of $l_1$ and $l_2$ regularization, 
which receives input of the two recent bids of the player (``lag-2 input''); we 
%The input to this model is the two recent bids. We call this type of input ``lag-2 input'' and 
refer to this model as AR2; As a non-linear benchmark we use a random forest model with lag-2 input (RF2); As a deep-learning benchmark we use  multi-layered perceptron models with lag-2 input (MLP2); Facebook's Prophet model \cite{ taylor2018forecasting}:\footnote{See also: \url{https://research.fb.com/blog/2017/02/prophet-forecasting-at-scale/}.} 
 an additive regression model that is designed to produce smooth forecasts of scalar data across time and to capture long and short term trends, as well as periodic signals, and is a natural state-of-the-art benchmark for the series bid prediction task. Prophet is trained for each player on the full sequence of training bids and the corresponding date and hour in day of each bid. %For the stepahead prediction task we re-train Prophet before each prediction with the sequence of true bids up to the previous timestep. 
See more details on the implementation of 
these ML algorithms in Appendix \ref{sec:ml_methods}. %in the appendix.

\subsection{ML vs. Econometric-Based Methods in the In-Sample Prediction Setting} \label{subsec:insample_results}
%\FloatBarrier

\begin{figure*}[t]
\centering
\begin{subfigure}{.32\linewidth}
\includegraphics[width=1.00\linewidth]{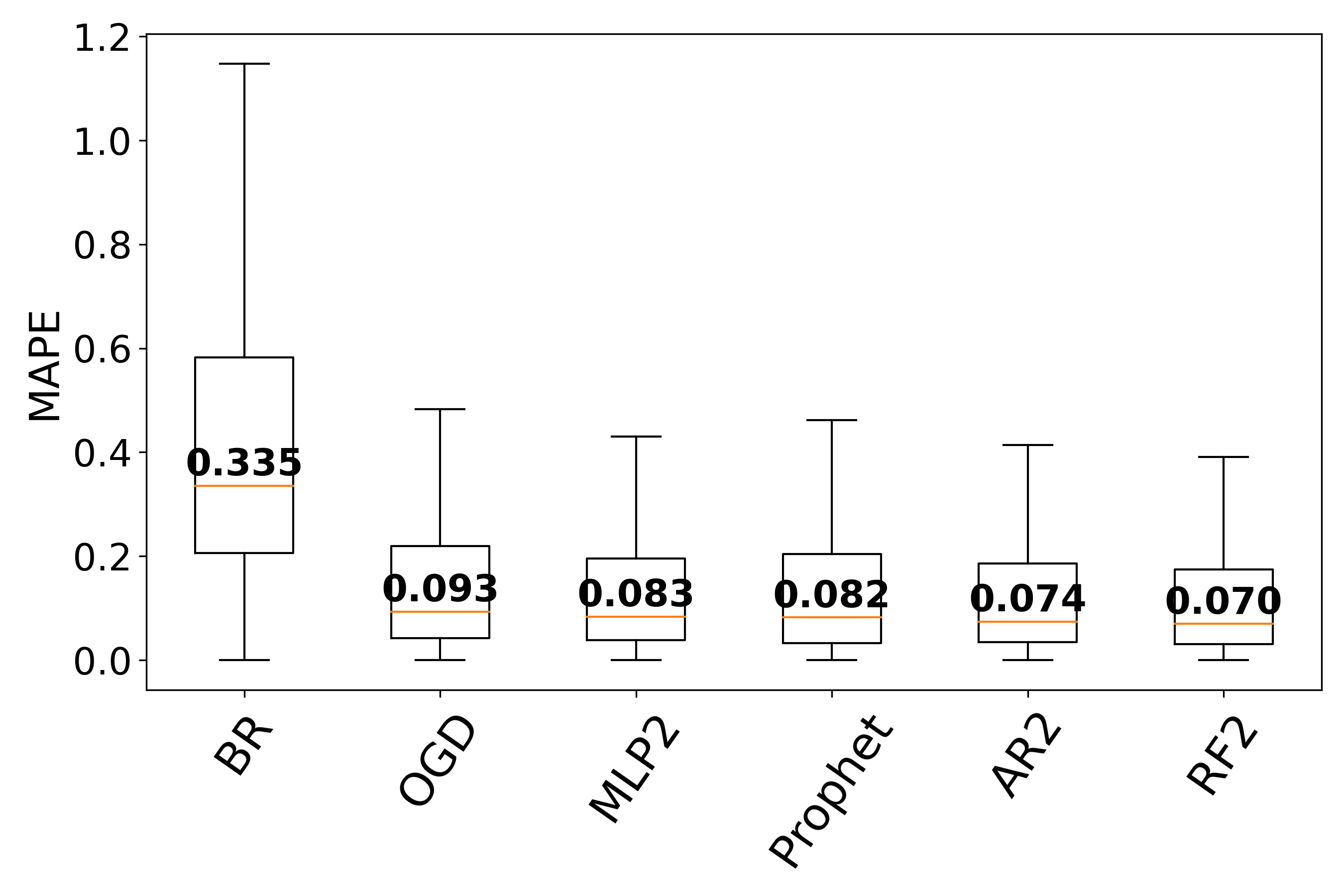}
\caption{Series Prediction.}  
\label{fig:insample_series_mape}
\end{subfigure}
\begin{subfigure}{.32\linewidth}
\includegraphics[width=1.00\linewidth]{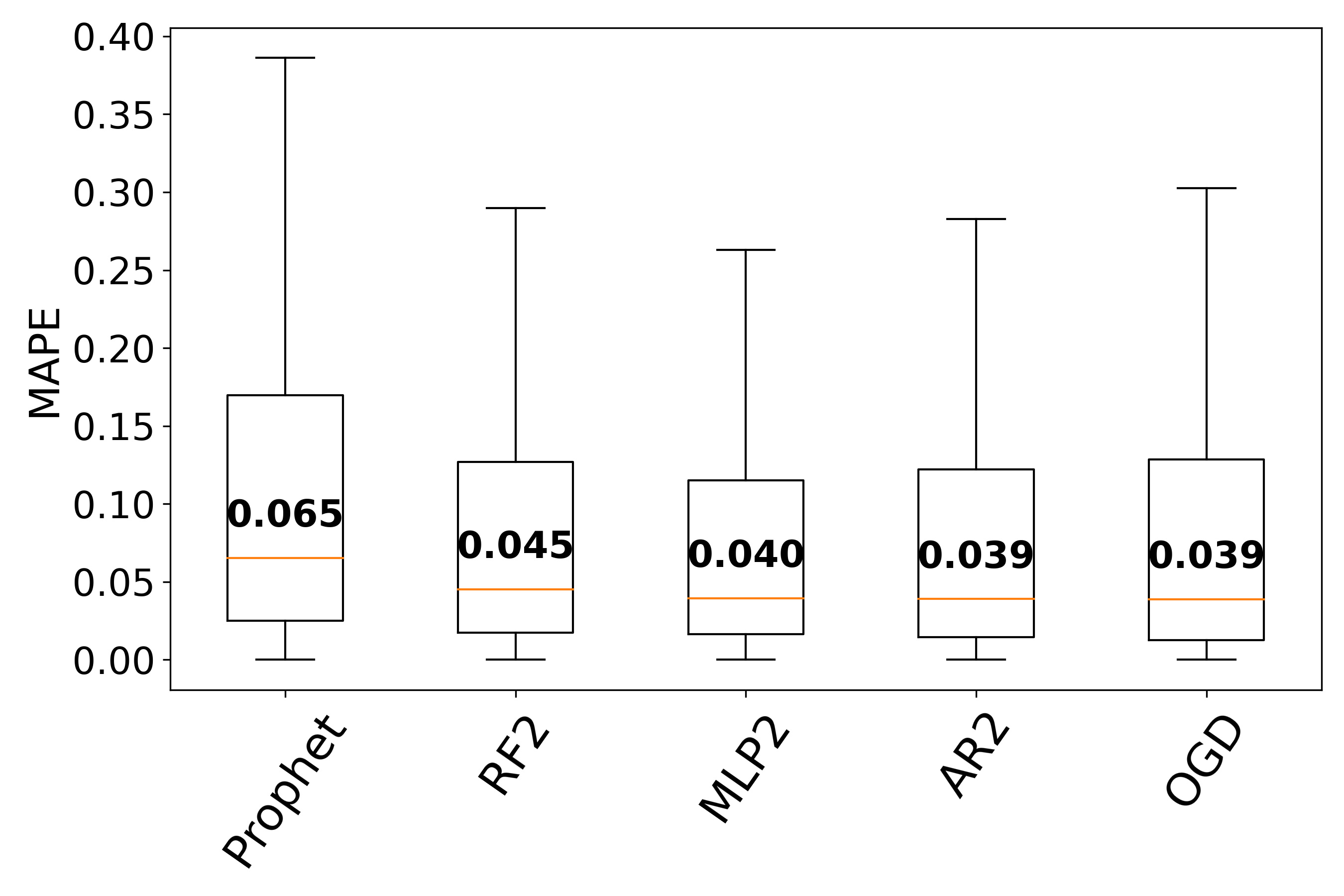}
\caption{Stepahead Prediction.}  
\label{fig:insample_stepahead_mape}
\end{subfigure}
\begin{subfigure}{.32\linewidth}
\includegraphics[width=1.00\linewidth]{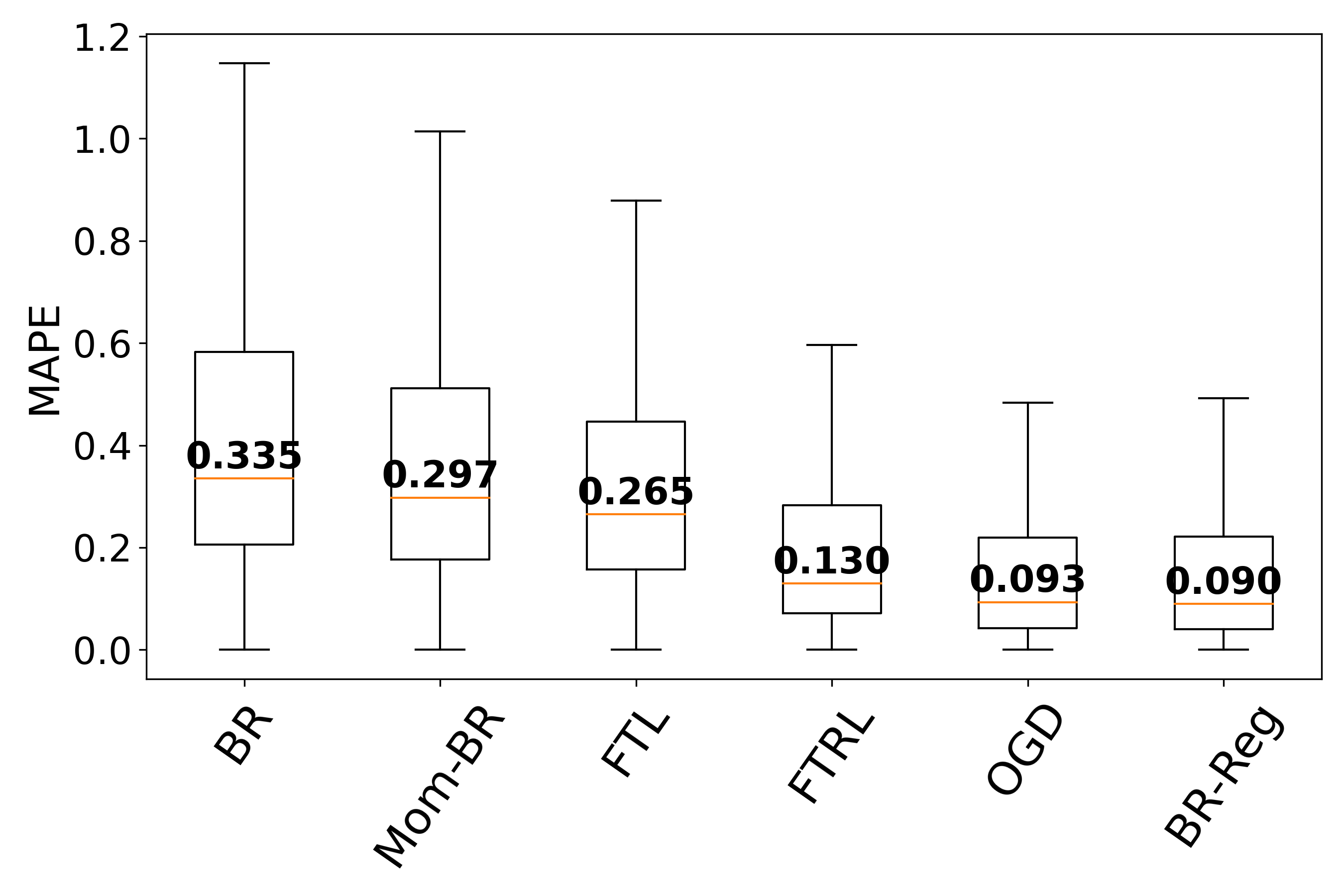}
\caption{Econ Methods (Series).}  
\label{fig:insample_econ_MAPE}
\end{subfigure}
\caption{Prediction performance in the in-sample setting (red lines denote the median MAPE). }
\label{fig:insample_mape}
\end{figure*}

\begin{table}[t]
\center
\scriptsize
\begin{subfigure}{.49\linewidth}
\centering
\begin{tabular}{lrrrr}
\toprule
{} &  mean &  stderr &    lb &    ub \\
\midrule
RF2            & 0.104 &   0.002 & 0.101 & 0.108 \\
AR2            & 0.107 &   0.002 & 0.104 & 0.111 \\
Prophet        & 0.114 &   0.002 & 0.111 & 0.118 \\
MLP2           & 0.117 &   0.002 & 0.114 & 0.121 \\
BR-Reg       & 0.118 &   0.002 & 0.115 & 0.122 \\
OGD            & 0.120 &   0.002 & 0.117 & 0.124 \\
FTRL & 0.152 &   0.002 & 0.149 & 0.156 \\
FTL            & 0.238 &   0.002 & 0.234 & 0.243 \\
Momentum-BR     & 0.253 &   0.002 & 0.249 & 0.257 \\
BR             & 0.274 &   0.002 & 0.270 & 0.279 \\
\bottomrule
\end{tabular}
\caption{Series prediction \label{fig:table_insample_series}}
\end{subfigure} \\
\begin{subfigure}{.49\linewidth}
\centering
\begin{tabular}{lrrrr}
\toprule
{} &  mean &  stderr &    lb &    ub \\
\midrule
MLP2 & 0.080 &   0.002 & 0.077 & 0.083 \\
AR2  & 0.080 &   0.002 & 0.077 & 0.083 \\
OGD       & 0.080 &   0.002 & 0.077 & 0.083 \\
RF2  & 0.085 &   0.002 & 0.081 & 0.088 \\
Prophet   & 0.101 &   0.002 & 0.098 & 0.104 \\
\bottomrule
\end{tabular}
\caption{Stepahead prediction \label{fig:table_insample_stepahead}}
\end{subfigure}
\caption{Mean MAPE score across players, standard error of the mean and 95\% confidence interval %, excluding outliers according to the standard method of  \cite{Tukey}. %(as indicated by bottom and top whiskers in the boxplots).
in the in-sample prediction setting. 
\label{fig:table_insample}}
\end{table}

Here we compare the predictive performance of the econometric-based methods and the ML benchmarks. We evaluate the prediction success by the Mean Absolute Percentage Error (MAPE) across bidders. The MAPE is defined for each bidder $i$ by $\frac{1}{|Test_i|}\sum_{t \in Test_i} |b^i_t-\hat{b}^i_t|/b^i_t$. We use percentage error since it naturally allows for aggregation of errors across bidders with bids of different scale, as the bids in our dataset (see Figure \ref{fig:bid_dist}). The MAPE distributions are presented in standard box plots, in which the boxes extend from the first to the third quartile values of the distribution, with a red line at the median. The whiskers are in the standard definition according to \cite{Tukey}. In addition, we provide for each figure a corresponding table with the mean errors and confidence intervals when excluding outliers according to %the standard method of 
\cite{Tukey}.

Figures \ref{fig:insample_series_mape} and \ref{fig:insample_stepahead_mape} show the MAPE distributions in the in-sample setting. %, in both the series and stepahead prediction tasks. 
Overall, the results show that in this setting, where the train and test sequences come from similar distributions, the bid-based ML methods %indeed 
do well. 
OGD  manages  to  achieve  comparable  results to the ML algorithms in terms of the main mass of the distributions in both the series and the stepahead tasks,  but  has higher  error  in   series prediction in terms of the mean error; this difference is statistically significant from RF2 and AR2, see Table \ref{fig:table_insample_series} 
for the mean errors and confidence intervals. 
We also see that BR is significantly inferior to the other methods. Note that BR is depicted only in the series prediction results; the predictions of BR are the same for the series and the stepahead prediction settings as they are not a function of the previous bids but of the economic feedback. 

Among the ML methods, the top performing methods in the series prediction task (Figure \ref{fig:insample_series_mape} and Table 
\ref{fig:table_insample_series}), 
are the non-linear RF2 model and the linear AR2 model.  
They outperform the state-of-the-art Prophet, probably  due to the relatively small training data, on which they manage to train more effectively. 
In the in-sample stepahead prediction task (Figure  \ref{fig:insample_stepahead_mape}), all methods except Prophet perform similarly well, with a median absolute percentage error of less than 5\%. Prophet is less suited for receiving step-by-step input and has the  worst performance in this stepahead task. This disadvantage is statistically significant in terms of the means (see Table \ref{fig:table_insample_stepahead}).
%C.1b). 
%In the curves we can see it still attempts to apply wave-like dynamics. see appendix.

In Appendix \ref{subsec:curves_insample} we show examples of bid curves and predictions of the OGD and the ML methods in the series and the stepahead prediction tasks.
The series prediction is a hard task  
%Projecting several steps into the future is a hard task 
since errors %of a model are served as inputs to the next prediction steps and the errors 
may be accumulated with time.
The examples show how Prophet manages to capture the bid dynamics well even in non-trivial dynamics, and that also OGD usually manages to
capture the correct direction of bid change. %In this 7d example, it catches the increasing bid trend except for the drop in bids that is expressed only in a flattening of the increase in predicted bids. 
%In 7a it is interesting to see that both Prophet and OGD outline similar curves, not far from the true bids, although the models are very different and are relying on very different inputs. All models predict relatively smooth curves compared to the large hourly fluctuations observed in the bidding data.
The predictions of %the non-linear RF2 and MLP2 and the linear AR2 
RF2, MLP2 and AR2  
are qualitatively similar, all usually ``cut'' the bid curves somewhere close to their average.
%The better MAPE score of RF2 shows that it is usually closer to the average than the other two methods.  Figure \ref{fig:insample_samples_stepahead} shows prediction examples for the same bidders in the stepahead prediction task, 
See Appendix \ref{subsec:curves_insample} %D.1
for more details. 

All in all, the econometric-based OGD achieves comparable results to the ML benchmarks, showing that players' actions are consistent with the economic feedback, as is captured by the utility functions estimated from the data (see Section \ref{sec:data}). In Section \ref{sec:gen} we show that the econometric-based methods that rely on the economic feedback are particularly useful in a setting where there is a change in the bid distributions.

\subsection{BR vs. OGD in the In-Sample Prediction Setting} \label{subsec:insample_econ}

%\begin{figure}[h!]
%\centerline{\includegraphics[scale=0%.3]{figures/final_boxplot_insample_e%con_no_aarp_mape}}
%\caption{Comparison of the %interpolating econ methods in the %in-sample setting.  %\label{fig:insample_econ_MAPE} }
%\end{figure}

We have seen that OGD predicts significantly better than BR. Figure \ref{fig:insample_econ_MAPE} shows the performance of the %methods that interpolate between BR and OGD, which 
econ methods that 
were presented in Section \ref{sec:interpolating_methods}. 
The figure also includes the Momentum-BR method, which is a direct interpolation between BR and OGD that sets the next bid to the average between the current bid and the best-reply bid. 
It can be clearly seen that the no-regret methods predict closer to the actual bidding data than the methods that do not minimize regret. 
FTL, which is a no-regret algorithm for strongly concave utility functions, has higher error than the classic no-regret learning algorithms. 
The figure also shows the effect of the two key features that distinguish between BR and OGD -- best reply to the history and regularization --  on the prediction performance.
A comparison of the memory-less BR and the FTL methods shows that replying to the entire history improves the performance compared to best-replying to the previous period only. However the more substantial improvement is obtained by adding a regularization term, as can be seen in the lower MAPE score of FTRL and BR-Reg.
Among the no-regret methods, the BR-Reg and the computationally-efficient OGD have the best performance, with an advantage that is statistically significant compared to the other methods (see Table \ref{fig:table_insample_series}).
%C.1a). 
%It makes sense that the OGD that is much more simple to rationalize and computationally efficient than the other two no-regret methods FTRL and BR-Reg, describes the bidders' behavior better.  

%\FloatBarrier

%\section{Predictive Performance with a Co-variate Shift}
\section{Predictive Performance in the Co-variate Shift Setting}
\label{sec:gen}

In this section we evaluate the methods in a setting where there is a change in the bid distribution. Our results show that in this more challenging {\em co-variate shift} prediction setting, 
there is a clear advantage to the econ-based  approach: 
the structured econ OGD method outperforms the unstructured bid-based ML benchmarks that fail to adapt to the change in the data.

\subsection{The Co-variate Shift Dataset}

\begin{figure}[t]
\centering
\begin{subfigure}{.49\linewidth}
\centering
\includegraphics[width=0.85\linewidth]{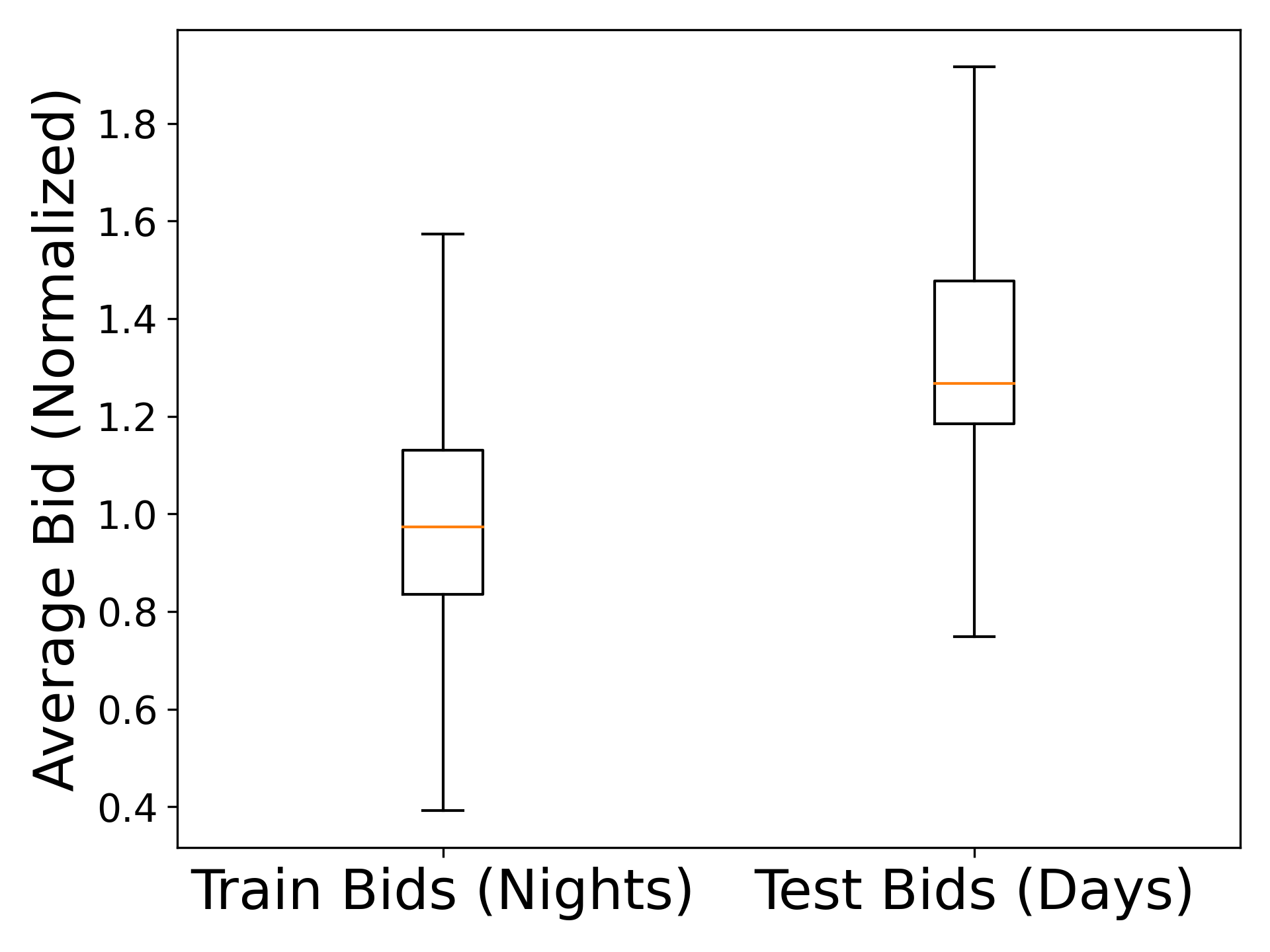}
\caption{Train-test bid distributions.}  
\label{fig:generalization_bid_dists}
\end{subfigure}
\begin{subfigure}{.49\linewidth}
\includegraphics[width=0.85\linewidth]{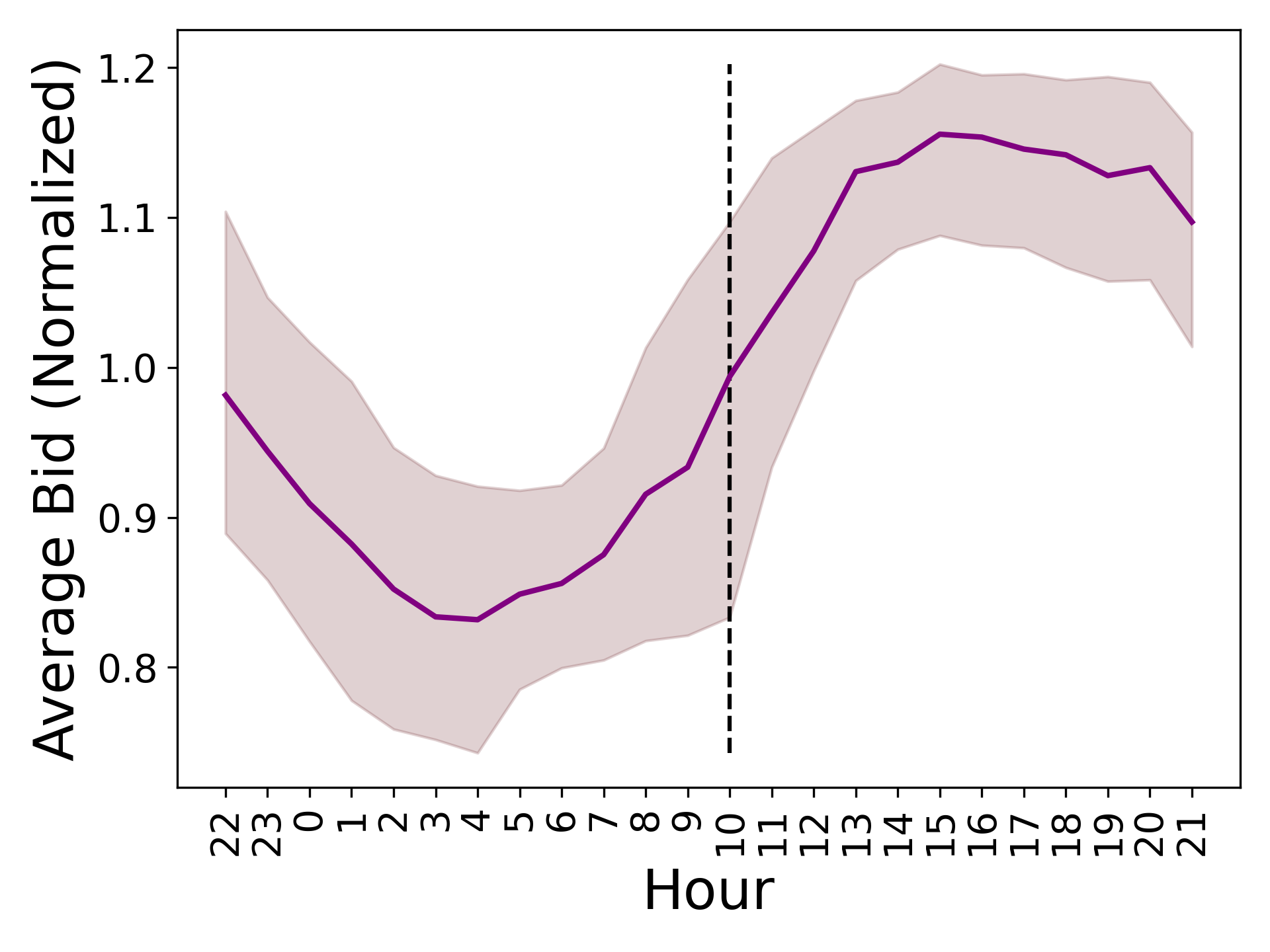}
\centering
\caption{Average bid in test days.}  
\label{fig:generalization_avg_bid}
\end{subfigure}
\caption{The co-variate shift dataset. }
\label{fig:generalization_dataset}
\end{figure}

We wish to evaluate the prediction performance in a setting in which the test bids are significantly different than the training bids. For this purpose, we subselect days from the data described in Section \ref{sec:data} (excluding the validation keyword), for which the distribution of bids during the day (10am to 9pm) is significantly different than the distribution of bids during the night (10pm to 9am), and use the day bids as the test set. Since the ML benchmarks usually benefit from larger training data than the 12 data points of a single night, we allow all methods to use all night data of a player as training. 
Specifically, to create the co-variate shift dataset, we consider full days with sufficiently non-trivial activity, i.e. days in which bidders participated in auctions throughout all 24 hours 
with a coefficient of variation of bids of at least 0.1. We subselect days where the day bids are different from all night bids for a player according to a Kolmogorov-Smirnov test (with $p < 0.001$) and apply a two-sided t-test to confirm that the day bid distribution is not only different from the train data but also has a statistically significant higher average (with $p < 0.05$).

In total, the co-variate shift dataset consists of %$816$ (with aarp)
$762$
days for prediction for %$279$ (with aarp)
$260$ 
bidders. The average number of training hours across bidders is %$142.3$ (with aarp) 
$141.2$
with a standard deviation of %29.6. (with aarp)
$29.8$. 
The number of test hours is $12$ for each test day (10am to 9pm), to a total of % $12 \cdot 861 = 10,332$  (with aarp)
$12 \cdot 762 = 9,144$
steps for prediction.		
Figure \ref{fig:generalization_bid_dists} shows the distributions of train bids (including all night bids for each player in the dataset) and test bids, both normalized by the average of the training bids for each player. The figure illustrates that indeed the distribution of test data is different than the training data. 
Figure \ref{fig:generalization_avg_bid} shows the average bid by the hour in day, across all test days in the co-variate shift dataset. The plot shows the average of test sequences on the right of the vertical line, and their preceding nights on the left, where each 24-hour sequence is normalized by its average. 
The shaded area shows the 25 to 75 percentiles for every hour. As can be seen, night bids are on average as low as 85\% of the average bid, and the test bids are on average as high as 115\% of the average bid.

\subsection{ML vs. Econometric-Based Methods in the Co-variate Shift Setting}

\begin{figure*}[t]
\centering
\begin{subfigure}{.32\linewidth}
\includegraphics[width=1.00\linewidth]{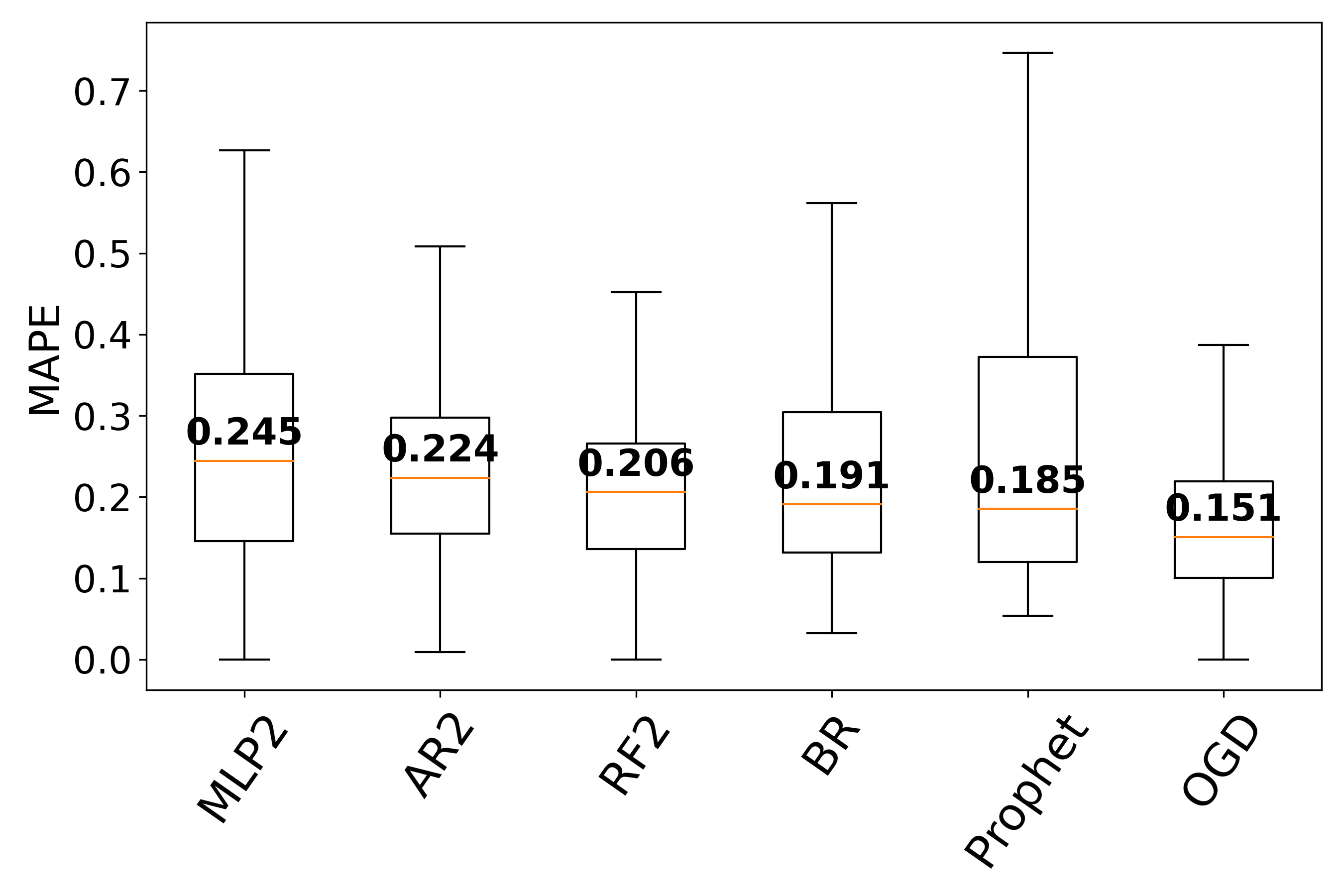}
\caption{Series Prediction.}  
\label{fig:generalization_series_mape}
\end{subfigure}
\begin{subfigure}{.32\linewidth}
\includegraphics[width=1.00\linewidth]{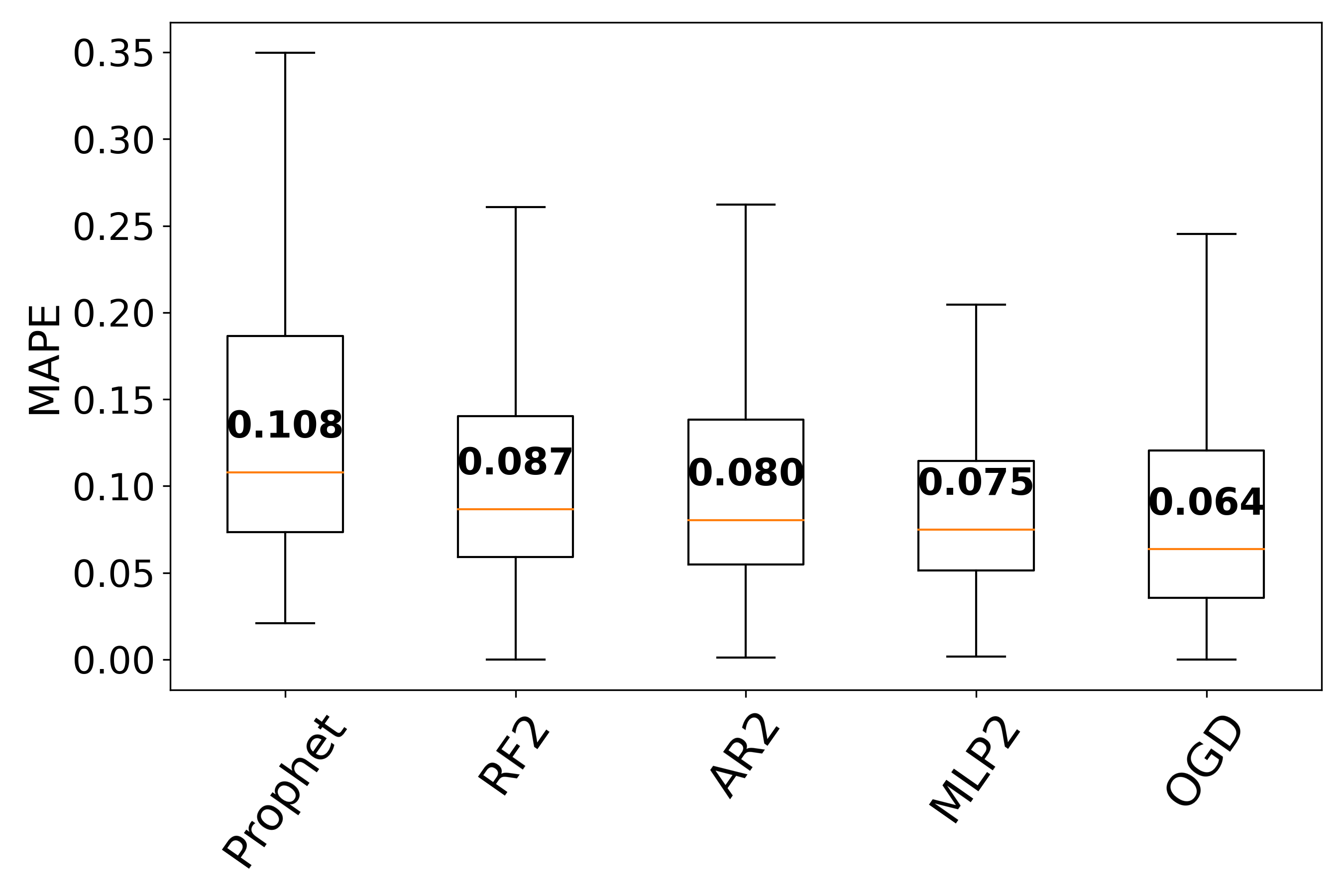}
\caption{Stepahead Prediction.}  
\label{fig:generalization_stepahead_mape}
\end{subfigure}
\begin{subfigure}{.32\linewidth}
\includegraphics[width=1.00\linewidth]{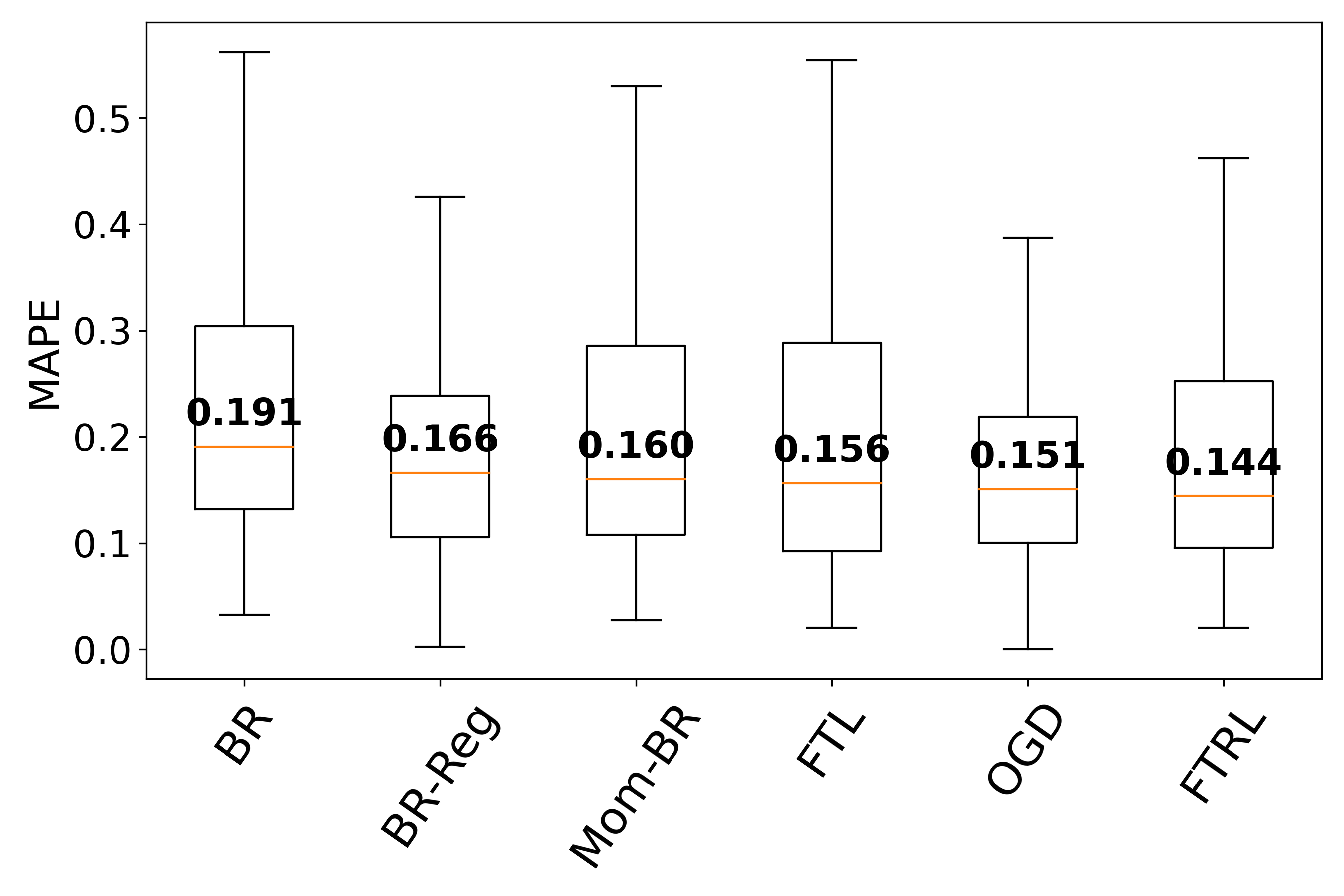}
\caption{Econ Methods (Series).}  
\label{fig:generalization_econ_MAPE}
\end{subfigure}
\caption{Prediction performance in the co-variate shift setting  (red lines denote the median MAPE). }
\label{fig:generalization_mape}
\end{figure*}

\begin{table}
\center
\scriptsize
\begin{subfigure}{.49\linewidth}
\centering
\begin{tabular}{lrrrr}
\toprule
{} &  mean &  stderr &    lb &    ub \\
\midrule
OGD            & 0.165 &   0.006 & 0.154 & 0.177 \\
FTL            & 0.168 &   0.007 & 0.155 & 0.181 \\
FTRL & 0.172 &   0.006 & 0.159 & 0.184 \\
BR-Reg       & 0.173 &   0.006 & 0.161 & 0.184 \\
Momentum-BR     & 0.177 &   0.006 & 0.164 & 0.190 \\
Prophet        & 0.191 &   0.007 & 0.178 & 0.204 \\
BR             & 0.197 &   0.006 & 0.185 & 0.210 \\
RF2            & 0.203 &   0.006 & 0.190 & 0.215 \\
AR2            & 0.221 &   0.006 & 0.209 & 0.233 \\
MLP2           & 0.226 &   0.007 & 0.212 & 0.241 \\
\bottomrule
\end{tabular}
\caption{Series prediction  \label{fig:table_gen_series}}
\end{subfigure} \\
\begin{subfigure}{.49\linewidth}
\centering
\begin{tabular}{lrrrr}
\toprule
{} &  mean &  stderr &    lb &    ub \\
\midrule
OGD       & 0.086 &   0.005 & 0.077 & 0.095 \\
MLP2 & 0.092 &   0.004 & 0.083 & 0.100 \\
AR2  & 0.104 &   0.005 & 0.094 & 0.113 \\
RF2  & 0.106 &   0.005 & 0.097 & 0.114 \\
Prophet   & 0.133 &   0.006 & 0.122 & 0.144 \\
\bottomrule
\end{tabular}
\caption{Stepahead prediction}
\end{subfigure}
\caption{Mean MAPE score across players, standard error of the mean and 95\% confidence interval %, excluding outliers according to the standard method of  \cite{Tukey}. 
in the co-variate shift prediction setting. 
\label{fig:table_gen}}
\end{table}

%Here we compare the prediction performance of the bid-based ML methods with the structured econ methods BR and OGD in the co-variate shift setting. %As in the previous section, we present the performance of the methods by their MAPE distribution.
Figures \ref{fig:generalization_series_mape} and \ref{fig:generalization_stepahead_mape} show the MAPE distributions of the bid-based ML methods and the structured-econ methods BR and OGD in the co-variate shift setting. 
Table \ref{fig:table_gen} % C.2 
presents the mean MAPE %scores 
across players as well as the confidence bounds.
Clearly, the simple and computationally-efficient OGD that relies on economic feedback, achieves the best performance both in the series and in the stepahead prediction tasks. The difference is statistically significant in all comparisons of OGD with the ML methods, except for the comparison with MLP2 and AR2 in the stepahead task. 
%The bid-based ML methods have a higher error, and even 
Also the BR method, that had the worst performance in the in-sample setting (see Section \ref{sec:insample}) is now  comparable %and even slightly better than 
to the bid-based ML methods. These results show that when there is a change in the data, it is better to react to the economic feedback than to the bid history that is no longer relevant. 

In the series prediction task (Figure ~\ref{fig:generalization_series_mape} and Table \ref{fig:table_gen_series}), among the ML methods, Prophet has the best median error, but still makes large errors for some of the players, as seen by the distribution width. RF2 has the second-best median, followed by AR2 and MLP2. %, all having more stable prediction performance across bidders than the Prophet model. 
Note that in this co-variate shift setting the length of the predicted series is $12$, while in the in-sample setting (Section \ref{sec:insample}) the average predicted series length is $20$, and therefore the error levels are not comparable. %as longer series predictions may accumulate larger errors. 
In contrast, it is possible to compare error levels across settings in the stepahead prediction. 
Figure \ref{fig:generalization_stepahead_mape} 
shows that as could be expected, in the challenging co-variate prediction setting all methods have higher median errors than in the in-sample setting. The OGD has the best performance with a median absolute error of $6.4$\% of the true bid. %As in the in-sample setting. Prophet that is not designed for a step-by-step update, has the worst performance with a median absolute error of $10.3$\% of the true bid. %Also the RF2 switches places and has the second-worst performance of $8.6$\%, and the best performance among the ML methods is of MLP2, with a median error of $6.9$\% of the true bid.

To portray the qualitative difference of the different prediction methods,  we present in the Appendix the prediction curves of the OGD and the ML methods alongside the actual bids, for a sample of bidders. The plots demonstrate how OGD typically matches the new (higher) bid level of the test data. In contrast, the ML methods fail to adapt to the new level of bids and their predictions typically remain close to the lower bids on which they were trained. See Section \ref{subsec:curves_generalization} in the Appendix 
%Appendix D.2 
for more details.

Figure \ref{fig:generalization_econ_MAPE} shows the performance of the econ methods %interpolating between BR and OGD 
that were discussed in Section  \ref{sec:interpolating_methods}. Unlike the results in the in-sample setting (see Section \ref{subsec:insample_results}), in the co-variate shift setting the differences between the econ methods are smaller and 
the separation between no-regret methods and the methods that do not minimize regret is less clear. However, the results still show that the BR method, which does not incorporate any element of learning, is inferior to the other methods that do combine some form of %no-regret
learning, with a difference that is statistically significant (see Table \ref{fig:table_gen_series}).

\section{ML Models with Economic Features} \label{sec:econ_ml}

\begin{figure}[t]
\centering
\begin{subfigure}{.49\linewidth}
\includegraphics[width=0.90\linewidth]{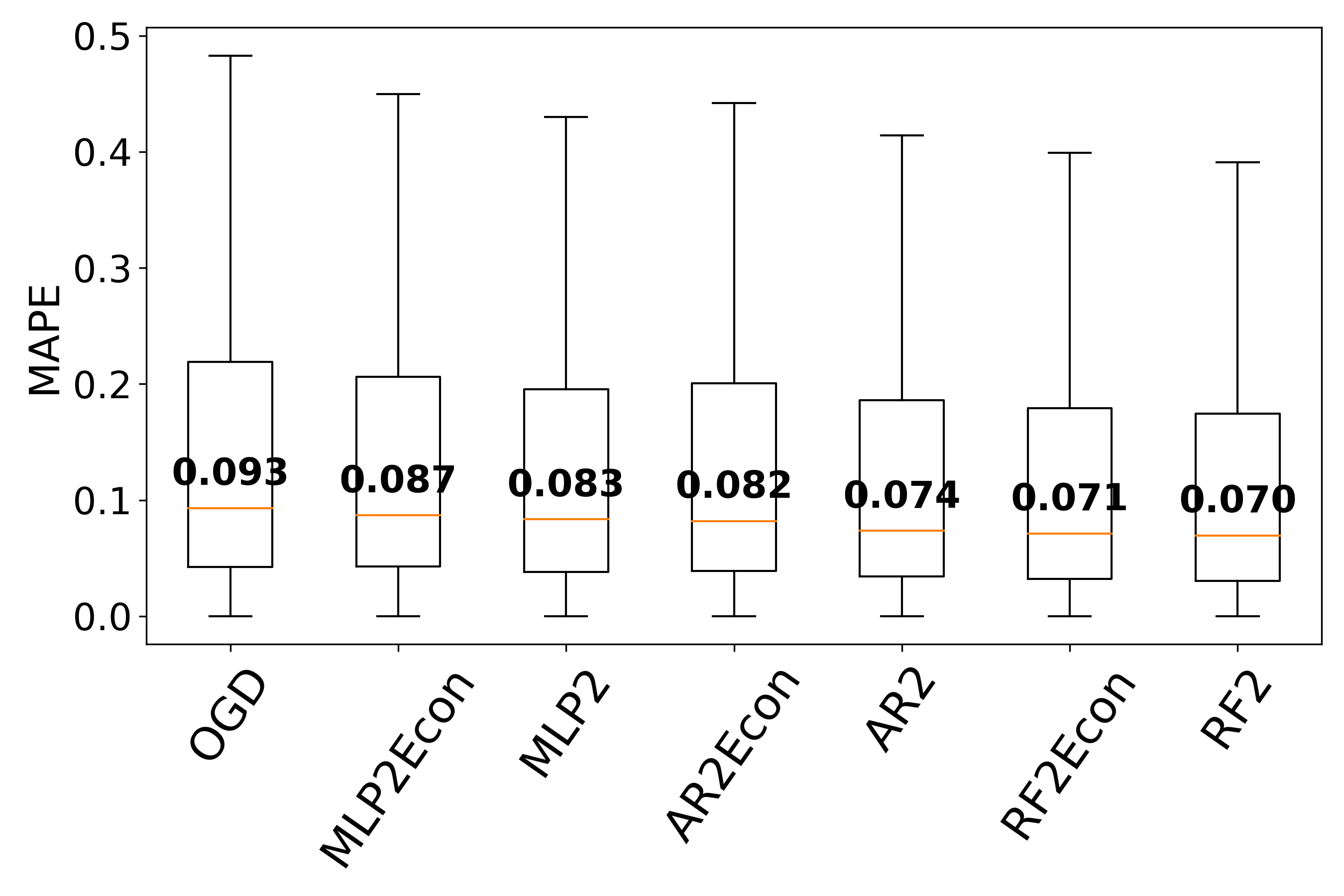}
\caption{In-sample Prediction.}  %In-sample Prediction Setting
\label{fig:insample_ml_econ}
\end{subfigure}
\begin{subfigure}{.49\linewidth}
\includegraphics[width=0.90\linewidth]{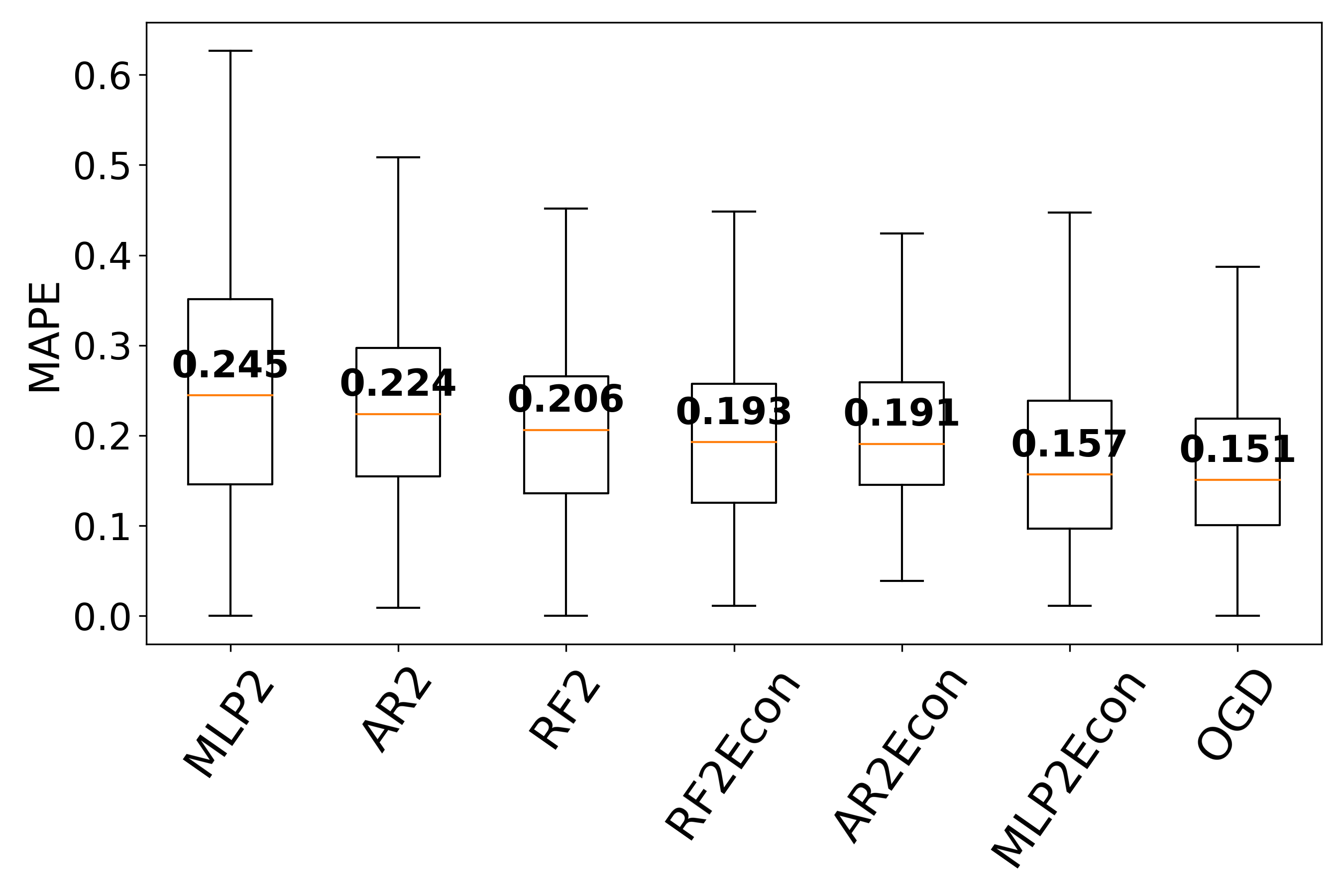}
\caption{Co-Variate Shift Prediction.}  %Co-Variate Shift Prediction Setting
\label{fig:generalization_ml_econ}
\end{subfigure}
\caption{Series prediction performance of the ML methods with economic features (red lines denote the median MAPE)}
\label{fig:ml_econ}
\end{figure}

We have seen that in the in-sample prediction setting the bid-based ML methods do well, while in the co-variate shift prediction setting they fail to adapt to the new bid distributions. 
Here we ask whether adding economic feedback to the bid-based input of the ML methods can improve their performance. 
Specifically, in addition to the two recent predicted\footnote{As explained previously, at the series prediction task the methods do not know anything about the bids in the test set, and rely at every step on preceding predictions that may potentially accumulate error.} bids (the ``lag-2 input''), the ML algorithms now receive as input the click and cost curves at the two recent predicted bids, as well as the gradients of the click and cost curves at the previous predicted bid. 

Figure \ref{fig:ml_econ} shows the MAPE results for series prediction in the in-sample and the co-variate shift prediction settings. 
In the in-sample setting (Figure \ref{fig:insample_ml_econ}), it can be seen that the models with the additional economic features lead to similar or slightly worse performance than the purely bid-based models. Thus, in this prediction setting, the bid information was a sufficient predictor of future bids and the extra econ features only introduced noise to the prediction. This is consistent with recent results in repeated normal-form games \cite{ijcai2019}.
In contrast, Figure \ref{fig:generalization_ml_econ} shows that in the co-variate shift prediction setting in which the bids in the training data have different distribution than the test data, the economic features can be useful in augmenting the ML methods.
The most significant utilization of the economic features is achieved by the MLP2 method; 
MLP2 with economic features (``MLP2Econ'') outperforms all other ML methods, with error that is only slightly higher than the OGD, 
while the bid-based MLP2 (i.e., the network model that did not receive the additional economic features) has the worst performance in this setting. 
Still, the simple econometric-based OGD method that models regret-minimizing players has the best performance.
These results further highlight the importance of economic feedback when there are changes in the market that lead to changes in bid distributions, and the usefulness of structural econometric methods in this setting.

%\section{Further Improvements for OGD with Insight on Advertiser Behavior}
\section{Further Improvements for OGD with Behavioral Insights}

%So far we have seen the usefulness and importance of using structural econometric models that assume no-regret learners and rely on economic information in predicting future bids.
So far we have seen that structural econometric models that assume no-regret learners and rely on economic information achieve comparable performance to state-of-the-art bid-based machine-learning benchmarks in the in-sample setting (Section \ref{sec:insample}), and outperform these benchmarks in the co-variate shift prediction setting (Section \ref{sec:gen}). 
This opens new questions as to what should this structural model be. In this section we demonstrate that it is possible to incorporate insights on bidders' behavior in the structure of these models to further improve their prediction performance. 

We focus on the objective that the no-regret learners are optimizing. In the analysis we presented so far, the basic assumption is that players are maximizing a quasi-linear utility function, which is the typical assumption in game-theoretic models. %That is, the analysis so far assumed that the utility function of the player is her value for the auction outcome minus her payment for this outcome. 
Now we consider OGD players who have a bias towards a certain visibility level. 
We show that imposing structure that takes this  {\em visibility bias} into account improves the bid prediction performance of OGD. %, thus suggesting that the biased utility function provides a more accurate modeling of advertiser behavior. 

Formally, as in previous sections, consider a bidder who participates in a sequence of sponsored search auctions, with a value-per-click $v$, and at each time $t$ has a concave and bounded click curve $x_t: \R_+\to \R_+$ and a cost-per-click (CPC) curve $p_t: \R_+\to \R_+$.
Thus, the quasi-linear utility function of the player at $t$ takes the form: 
$\tilde{u}_t(b;v) = (v - p_t(b))x_t(b)$. 
We say that a bidder with a visibility bias perceives an additional utility term that takes into account her distance from a certain target visibility level. Specifically, let $xmax_t=\lim_{b \rightarrow \infty} x_t(b)$ denote the supremum click rate of the player at time $t$.\footnote{Note that it is not necessarily that an infinity bid would grant the bidder the first slot, as beyond the bid there may be additional considerations of the auction platform that determine the threshold for the first slots.} The visibility level of bid $b$ for the bidder at time $t$ is defined as $vis_t(b) = \frac{x_t(b)}{xmax_t}$. Let $vis_0 \in [0,1]$ denote the particular visibility level that the bidder is targeting. Similar to the value $v$ of the bidder, we assume that her target visibility level remains constant throughout the auction sequence that is being considered. 
Then, we model the visibility-based utility function that the player is optimizing as: 
\begin{align}
    u_t(b;v, \alpha, vis_0) = \tilde{u}_t(b;v) + \frac{1}{2}\alpha (vis_t(b) - vis_0)^2
\end{align}
  
 where the bidder's visibility coefficient $\alpha$ is the strength of her visibility bias. %Note that in order to compare the strength of the bias between players, $\alpha$ has to be rescaled according to the typical magnitude of the quasi-linear utility of the player (e.g., by the average utility of the player), or to consider the ratio between the bias term and the quasi-linear utility term. 

We evaluate the bid-prediction performance of the OGD model with the visibility-biased players versus the baseline OGD model in which bidders optimize the standard quasi-linear utility function. An OGD player with visibility bias plays according to the following update rule (\ogdbehav): 
\begin{align}
    b_{t+1} = h_{\ogdbehav, \eta, v, \alpha, vis_0}(b_t, u_t) := b_t + \eta \nabla_{b} u_t(b_t; v, \alpha, vis_0)\\
		= h_{\ogd, \eta, v}(b_t, \tilde{u}_t) + \eta \nabla_{b} \left(\frac{1}{2}\alpha (vis_t(b) - vis_0)^2\right)
\end{align}

The two additional parameters of \ogdbehav, namely, $vis_0$ and $\alpha$, are fitted for each player via direct grid optimization of the MAPE score on the training data. For $\alpha$ we searched on a grid in $[0,300]$ and for $vis_0$ a grid in $[0,1]$. 
Figure \ref{fig:params} in the Appendix shows the histogram of the parameters obtained from the optimization, across players.

\begin{figure}[t]
\centering
\begin{subfigure}{.49\linewidth}
\includegraphics[width=0.90\linewidth]{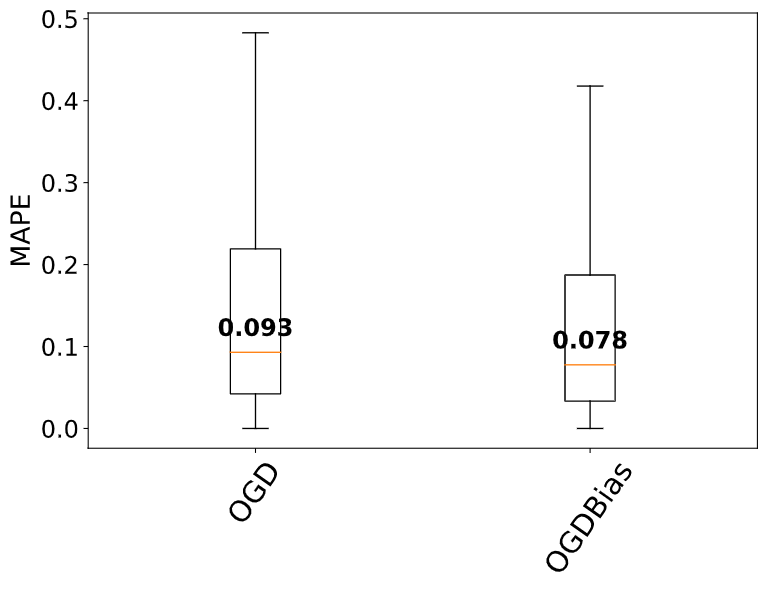}
\centering
\caption{}  
\label{fig:boxplot}
\end{subfigure}
\centering
\begin{subfigure}{.49\linewidth}
\includegraphics[width=1.00\linewidth]{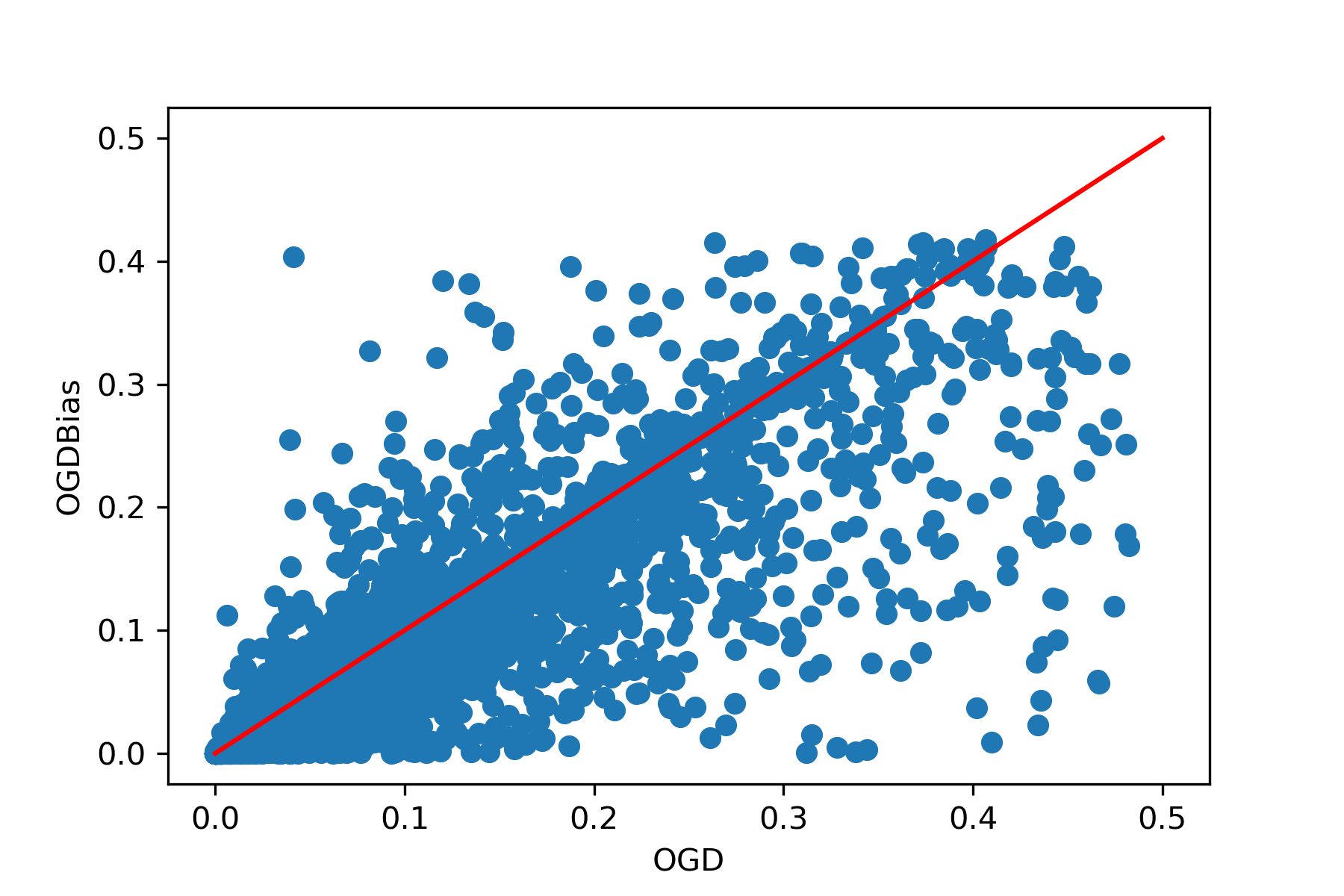}
\vspace{5pt}
\caption{}  
\label{fig:scatter}
\end{subfigure}
\caption{MAPE Distribution: \ogdbehav vs. OGD prediction performance.}
\label{fig:mape}
\end{figure}

Figure \ref{fig:mape} compares the prediction performance of  \ogdbehav  with the OGD  method that assumes the standard quasi-linear utility in the in-sample prediction setting. Figure \ref{fig:boxplot} shows the MAPE distributions of the two methods. 
As can be seen, \ogdbehav reduced the median MAPE across players by more than $15\%$, from $0.093$ to $0.078$. The mean MAPE (excluding outliers as in Table~\ref{fig:table_insample}) of the \ogdbehav method and the OGD method is $0.108\pm 0.004$ and $0.120 \pm 0.004$, respectively, and the difference is statistically significant. Figure \ref{fig:scatter} depicts the MAPE of \ogdbehav vs. the MAPE of OGD for all bidders and shows that the vast majority of the players are below the diagonal $y=x$. That is, for the vast majority of the players, the visibility-biased utility function provides a more accurate modeling of players' bidding dynamics than the standard quasi-linear utility function when considering the OGD update rule. In particular, we find that for 68\% of the bidders, \ogdbehav improves performance compared to the standard OGD method, for 46\% of the bidders the improvement is greater than 10\% of the OGD MAPE, and for above 25\% of the bidders the improvement is greater than 30\% of the OGD MAPE. On the other hand only 13\% of the bidders had a deterioration of the MAPE of more than 10\%. Since the \ogdbehav model is a more general model than OGD, the latter primarily stems from overfitting and finite sample error introduced by the extra parameters of the \ogdbehav model. Our results though indicate that the benefits of the extra model parameters substantially outbalance the extra variance introduced by the larger number of parameters.
%, i.e., \ogdbehav predicted future bids more accurately than OGD for these players. This suggests that these players are more accurately modeled as visibility-biased players than with the standard utility function. 

% Figure \ref{fig:scatter} also shows that there is a variance in this improvement, and that for some of the bidders (above the diagonal) the more flexible \ogdbehav model leads to overfitting. 
% Figure \ref{fig:overfit} in the Appendix presents the %frequency of improvement and overfitting. 
% distributions of bidders for which \ogdbehav leads to an improvement and to overfitting compared to OGD.
% {\color{red} The figure shows that overfitting due to the \ogdbehav model is significantly less prevalent than the improvement it obtains. 
% For example, for about 70\% of the bidders,  \ogdbehav improves the performance compared to the standard OGD method, for almost half of the bidders the improvement is greater than 10\% of the OGD MAPE, and for above 23\% of the bidders the improvement is greater than 30\% of the OGD MAPE.}

\section{Conclusions}

This work empirically evaluates the usefulness of econometric methods for learning agents for predicting future bidding behavior, on a large sponsored search bidding dataset from Microsoft's BingAds platform. 
Our dataset includes conterfactual data of the click rate and cost-per-click for each bidder. We propose how to use these economic fundamentals and econometric approaches to derive predictions for future bid dynamics in repeated auctions within the no-regret learning framework. Our empirical evaluation portrays the importance of the economic and econometric approach to bid prediction, and in particular the usefulness of the no-regret learning approach to predict bid dynamics in a changing market. 

More specifically, the evaluation we have presented shows that as long as there is no particular change in the data, the recent bids of a bidder are a good predictor of the bidder's future behavior. In this case, bid-based machine-learning methods are indeed useful for the prediction task, and the econometric-based methods achieve comparable results. However, when a co-variate shift arises in the data, these machine-learning methods fail to adapt to the change and result in high prediction errors. This change-in-circumstances setting stresses the usefulness of the structural econometric approach to bid prediction: our results show that econometric-based methods that rely on economic feedback outperform the machine-learning methods both in the task of predicting a series of future bids and in the task of predicting one step at every time. We demonstrate how in this setting the economic feedback can be useful also for augmenting machine-learning methods. Thus, once a change occurs, economic feedback is a more reliable predictor than past bids.
In addition, among the econometric-based methods we have found that no-regret learning methods outperform methods that are based on the traditional assumption that players best-respond to the competition at every step.

These results suggest various research directions regarding the use of economic and econometric theories for predicting bidding behavior. This includes questions that range from studying further the contribution of economic feedback for augmenting machine-learning methods, to studying the underlying economic modeling assumptions, such as the objective function and the optimization strategy that the bidders are using. We presented promising results in this vein, which demonstrate that an addition of a visibility bias component to the standard quasi-linear utility function of the bidder significantly improves the prediction performance. We believe these research directions can lead to further improvements in bid prediction and efficiency in auction markets, as well as to contribute to our understanding of bidder behavior in repeated auctions.

%\begin{verbatim} needed?
%\begin{acks}
%...
%\end{acks}
%\end{verbatim}

\bibliographystyle{splncs03}
\bibliography{references}

\begin{thebibliography}{10}
\providecommand{\url}[1]{\texttt{#1}}
\providecommand{\urlprefix}{URL }

\bibitem{Adamskiy2012}
Adamskiy, D., Koolen, W.M., Chernov, A., Vovk, V.: A closer look at adaptive
  regret. In: Proceedings of the 23rd International Conference on Algorithmic
  Learning Theory. pp. 290--304. ALT12, Springer-Verlag, Berlin, Heidelberg
  (2012), \url{https://doi.org/10.1007/978-3-642-34106-9_24}

\bibitem{Alaei2019}
Alaei, S., Badanidiyuru, A., Mahdian, M., Yazdanbod, S.: Response prediction
  for low-regret agents. In: Caragiannis, I., Mirrokni, V., Nikolova, E. (eds.)
  Web and Internet Economics. pp. 31--44. Springer International Publishing,
  Cham (2019)

\bibitem{athey2010structural}
Athey, S., Nekipelov, D.: A structural model of sponsored search advertising
  auctions. In: Sixth ad auctions workshop. vol.~15 (2010)

\bibitem{Genie2019}
Bayir, M.A., Xu, M., Zhu, Y., Shi, Y.: Genie: An open box counterfactual policy
  estimator for optimizing sponsored search marketplace. In: Proceedings of the
  Twelfth ACM International Conference on Web Search and Data Mining. pp.
  465--473. WSDM'19, Association for Computing Machinery, New York, NY, USA
  (2019), \url{https://doi.org/10.1145/3289600.3290969}

\bibitem{Box1976}
Box, G.E.P.: Science and statistics. Journal of the American Statistical
  Association  71(356),  791--799 (1976),
  \url{https://www.tandfonline.com/doi/abs/10.1080/01621459.1976.10480949}

\bibitem{Braverman2018}
Braverman, M., Mao, J., Schneider, J., Weinberg, M.: Selling to a no-regret
  buyer. In: Proceedings of the 2018 ACM Conference on Economics and
  Computation. pp. 523--538. EC'18, Association for Computing Machinery, New
  York, NY, USA (2018), \url{https://doi.org/10.1145/3219166.3219233}

\bibitem{caragiannis2015bounding}
Caragiannis, I., Kaklamanis, C., Kanellopoulos, P., Kyropoulou, M., Lucier, B.,
  Leme, R.P., Tardos, {\'E}.: Bounding the inefficiency of outcomes in
  generalized second price auctions. Journal of Economic Theory  156,  343--388
  (2015)

\bibitem{edelman2007internet}
Edelman, B., Ostrovsky, M., Schwarz, M.: Internet advertising and the
  generalized second-price auction: Selling billions of dollars worth of
  keywords. American economic review  97(1),  242--259 (2007)

\bibitem{Fudenberg2014}
Fudenberg, D., Levine, D.K.: Recency, consistent learning, and nash
  equilibrium. Proceedings of the National Academy of Sciences of the United
  States of America  111,  10826--10829 (2014),
  \url{http://www.jstor.org/stable/23800667}

\bibitem{Hazan2007strongly}
Hazan, E., Agarwal, A., Kale, S.: Logarithmic regret algorithms for online
  convex optimization. Mach. Learn.  69(2-3),  169--192 (dec 2007)

\bibitem{Hazan2009}
Hazan, E., Seshadhri, C.: Efficient learning algorithms for changing
  environments. In: Proceedings of the 26th Annual International Conference on
  Machine Learning. pp. 393--400. ICML'09, Association for Computing Machinery,
  New York, NY, USA (2009), \url{https://doi.org/10.1145/1553374.1553425}

\bibitem{meanfield2014}
Iyer, K., Johari, R., Sundararajan, M.: Mean field equilibria of dynamic
  auctions with learning. Management Science  60(12),  2949--2970 (2014)

\bibitem{kingma2014adam}
Kingma, D.P., Ba, J.: Adam: A method for stochastic optimization. arXiv
  preprint arXiv:1412.6980  (2014)

\bibitem{ijcai2019}
Kolumbus, Y., Noti, G.: Neural networks for predicting human interactions in
  repeated games. In: Proceedings of the Twenty-Eighth International Joint
  Conference on Artificial Intelligence, {IJCAI-19}. pp. 392--399.
  International Joint Conferences on Artificial Intelligence Organization (7
  2019), \url{https://doi.org/10.24963/ijcai.2019/56}

\bibitem{lewis2015unfavorable}
Lewis, R.A., Rao, J.M.: The unfavorable economics of measuring the returns to
  advertising. The Quarterly Journal of Economics  130(4),  1941--1973 (2015)

\bibitem{lucier2012revenue}
Lucier, B., Paes~Leme, R., Tardos, E.: On revenue in the generalized second
  price auction. In: Proceedings of the 21st international conference on World
  Wide Web. pp. 361--370 (2012)

\bibitem{Tukey}
McGill, R., Tukey, J.W., Larsen, W.A.: Variations of box plots. The American
  Statistician  32(1),  12--16 (1978)

\bibitem{Mohri2016}
Mohri, M., Medina, A.M.n.: Learning algorithms for second-price auctions with
  reserve. J. Mach. Learn. Res.  17(1),  2632--2656 (Jan 2016)

\bibitem{Nekipelov2015}
Nekipelov, D., Syrgkanis, V., Tardos, E.: Econometrics for learning agents. In:
  Proceedings of the Sixteenth ACM Conference on Economics and Computation. pp.
  1--18. EC'15, Association for Computing Machinery, New York, NY, USA (2015),
  \url{https://doi.org/10.1145/2764468.2764522}

\bibitem{Ng2000}
Ng, A.Y., Russell, S.J.: Algorithms for inverse reinforcement learning. In:
  Proceedings of the Seventeenth International Conference on Machine Learning.
  pp. 663--670. ICML'00, Morgan Kaufmann Publishers Inc., San Francisco, CA,
  USA (2000)

\bibitem{Nisan2017a}
Nisan, N., Noti, G.: An experimental evaluation of regret-based econometrics.
  In: Proceedings of the 26th International Conference on World Wide Web. pp.
  73--81. WWW'17, International World Wide Web Conferences Steering Committee,
  Republic and Canton of Geneva, CHE (2017),
  \url{https://doi.org/10.1145/3038912.3052621}

\bibitem{Nisan2017b}
Nisan, N., Noti, G.: A ``quantal regret'' method for structural econometrics in
  repeated games. In: Proceedings of the 2017 ACM Conference on Economics and
  Computation. p. 123. EC '17, Association for Computing Machinery, New York,
  NY, USA (2017), \url{https://doi.org/10.1145/3033274.3085111}

\bibitem{paarsch2006introduction}
Paarsch, H.J., Hong, H., et~al.: An introduction to the structural econometrics
  of auction data. MIT Press Books  1 (2006)

\bibitem{scikit-learn}
Pedregosa, F., Varoquaux, G., Gramfort, A., Michel, V., Thirion, B., Grisel,
  O., Blondel, M., Prettenhofer, P., Weiss, R., Dubourg, V., Vanderplas, J.,
  Passos, A., Cournapeau, D., Brucher, M., Perrot, M., Duchesnay, E.:
  Scikit-learn: Machine learning in {P}ython. Journal of Machine Learning
  Research  12,  2825--2830 (2011)

\bibitem{Jiang2016}
Rong, J., Qin, T., An, B., Liu, T.Y.: Modeling bounded rationality for
  sponsored search auctions. In: Proceedings of the Twenty-Second European
  Conference on Artificial Intelligence. pp. 515--523. ECAI'16, IOS Press, NLD
  (2016), \url{https://doi.org/10.3233/978-1-61499-672-9-515}

\bibitem{Russel1998}
Russell, S.: Learning agents for uncertain environments (extended abstract).
  In: Proceedings of the Eleventh Annual Conference on Computational Learning
  Theory. pp. 101--103. COLT'98, Association for Computing Machinery, New York,
  NY, USA (1998), \url{https://doi.org/10.1145/279943.279964}

\bibitem{shalev2012}
Shalev{-}Shwartz, S.: Online learning and online convex optimization.
  Foundations and Trends in Machine Learning  4(2),  107--194 (2012)

\bibitem{Syrgkanis2015}
Syrgkanis, V., Agarwal, A., Luo, H., Schapire, R.E.: Fast convergence of
  regularized learning in games. In: Proceedings of the 28th International
  Conference on Neural Information Processing Systems - Volume 2. pp.
  2989--2997. NIPS'15, MIT Press, Cambridge, MA, USA (2015)

\bibitem{taylor2018forecasting}
Taylor, S.J., Letham, B.: Forecasting at scale. The American Statistician
  72(1),  37--45 (2018)

\bibitem{toulis2017}
Toulis, P., Airoldi, E.M.: Asymptotic and finite-sample properties of
  estimators based on stochastic gradients. Ann. Statist.  45(4),  1694--1727
  (08 2017), \url{https://doi.org/10.1214/16-AOS1506}

\bibitem{yu2019multi}
Yu, L., Song, J., Ermon, S.: Multi-agent adversarial inverse reinforcement
  learning. In: International Conference on Machine Learning. pp. 7194--7201
  (2019)

\bibitem{Zinkevich2003}
Zinkevich, M.: Online convex programming and generalized infinitesimal gradient
  ascent. In: Proceedings of the 20th international conference on machine
  learning (icml-03). pp. 928--936 (2003)

\end{thebibliography}

\newpage

\appendix

\section*{APPENDICES}

\section{Hypothesis Testing for OGD} \label{app:qr_values}

\begin{figure}[ht]
\centering
\includegraphics[width=.5\linewidth]{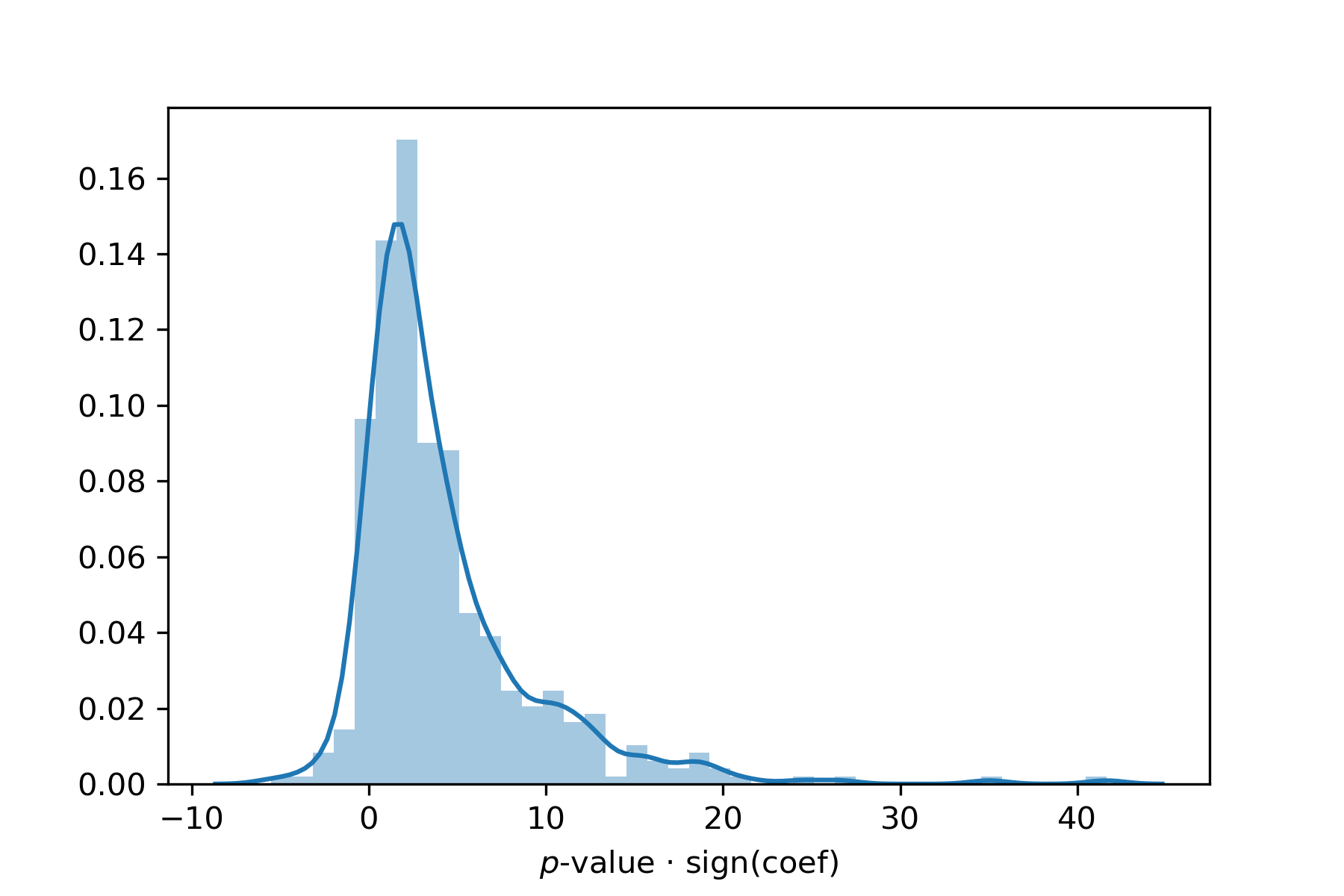}
\caption{Hypothesis testing related to the OGD algorithm's plausibility. Distribution across players of the negative log of $p$-value of the correlation between  $b_{t+1}-b_{t}$ and $\nabla_{b} u_t(b_t; v_{qr})$, multiplied by the sign of correlation. The distribution is shown for the subset of players with average $b_{t+1}-b_t$ of at least $1$ cent across their bid series. We observe that for almost all such players OGD with value $v_{qr}$ is a plausible model (as most signed $p$-values are strictly positive).}
\label{fig:grad_v_diff_ogd}
\end{figure}

%\section{Time-Series ML Methods}
\section{Implementation of the ML Benchmarks}

\label{sec:ml_methods}

We implemented the following machine learning (ML) benchmarks: %We used one keyword for hyperparameter tuning of the ML methods and then evaluated the methods on the remainder of the keywords.

\textbf{AR2:} A linear model that is implemented using LinearRegression in the python scikit-learn package \cite{scikit-learn}. The input to this model at each time step is the two recent bids. In the series prediction setting the two recent bids are the last two predictions made by the model and in the stepahead setting the input is the two recent true bids. We call this type of input ``lag-2 input.''
    
\textbf{RF2:} as a non-linear machine learning benchmark we use a random forest model with lag-2 input. The random forest predictor has 100 trees with a maximum depth of 2, and bootstrap sub-sampling was used to build each tree. The model is implemented using RandomForestRegressor in the scikit-learn python package.

\textbf{MLP2:} As a deep-learning benchmark we use multi-layered perceptron models (fully connected feed-forward neural networks) with lag-2 inputs. The networks have two hidden layers with 128 units in each layer, relu activation function and are optimized using the ADAM optimizer with the legacy parameters of \cite{ kingma2014adam} for 100 epochs, with a batch size of 10 and a learning rate of $0.0001$. The networks are implemented using scikit-learn MLPRegressor. We fitted the number of hidden layers, the size of each layer, the learning rate, and the number of ephocs on validation data of a single keyword that is separate from the test data analyzed in the paper.
    
\textbf{Prophet:} Facebook's Prophet model \cite{ taylor2018forecasting}\footnote{See also: \url{https://research.fb.com/blog/2017/02/prophet-forecasting-at-scale/}.}  is a modern additive regression model for time series forecasting. Prophet is trained for each player on the full sequence of training bids and the corresponding date and hour in day of each bid. Prophet is especially designed to produce smooth forecasts of scalar data across time and to capture long and short term trends, as well as periodic signals, and is a natural state-of-the-art benchmark for the series bid prediction task. For the stepahead prediction task we re-train Prophet before each prediction with the sequence of true bids up to the previous timestep and produce a new prediction at each step with the newly updated model. We implemented Prophet in python using the original source code available by the authors of \cite{ taylor2018forecasting}.

\FloatBarrier

\section{Distribution of Parameters for the OGDBias Method}

\FloatBarrier

\begin{figure}[h!]
\centering
\begin{subfigure}{.49\linewidth}
\includegraphics[width=0.85\linewidth]{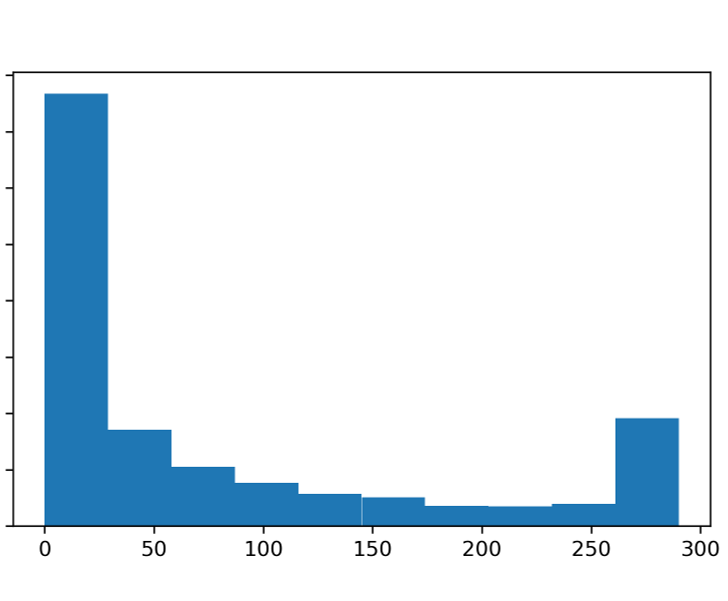}
\centering
\caption{Alpha}  
\label{fig:alpha}
\end{subfigure}
\centering
\begin{subfigure}{.49\linewidth}
\includegraphics[width=0.85\linewidth]{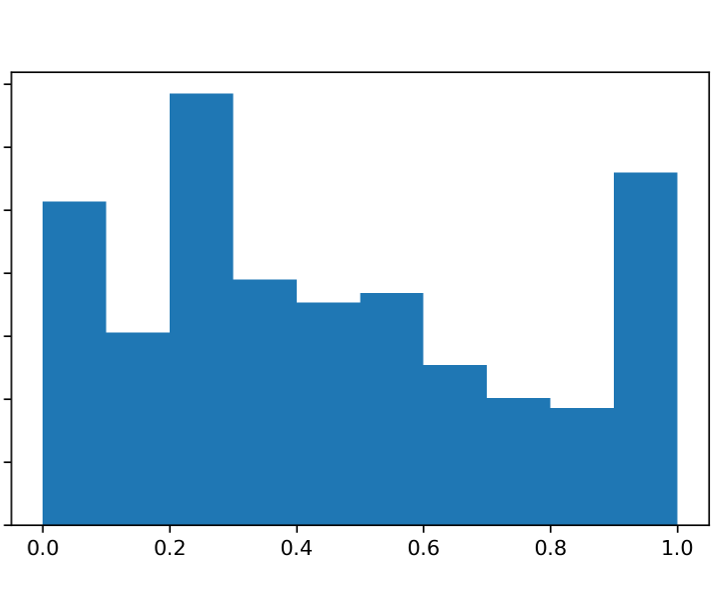}
\centering
\caption{$vis_0$}  
\label{fig:p0}
\end{subfigure}
\caption{Distribution of parameters.}
\label{fig:params}
\end{figure}

\FloatBarrier

\section{Prediction Examples}

%\subsection{Prediction Examples in the In-Sample Prediction Setting} \label{subsec:curves_insample}
\subsection{The In-Sample Prediction Setting} \label{subsec:curves_insample}

\begin{figure*}[t]
\centering
\begin{subfigure}{1.00\linewidth}
\includegraphics[width=1.00\linewidth]{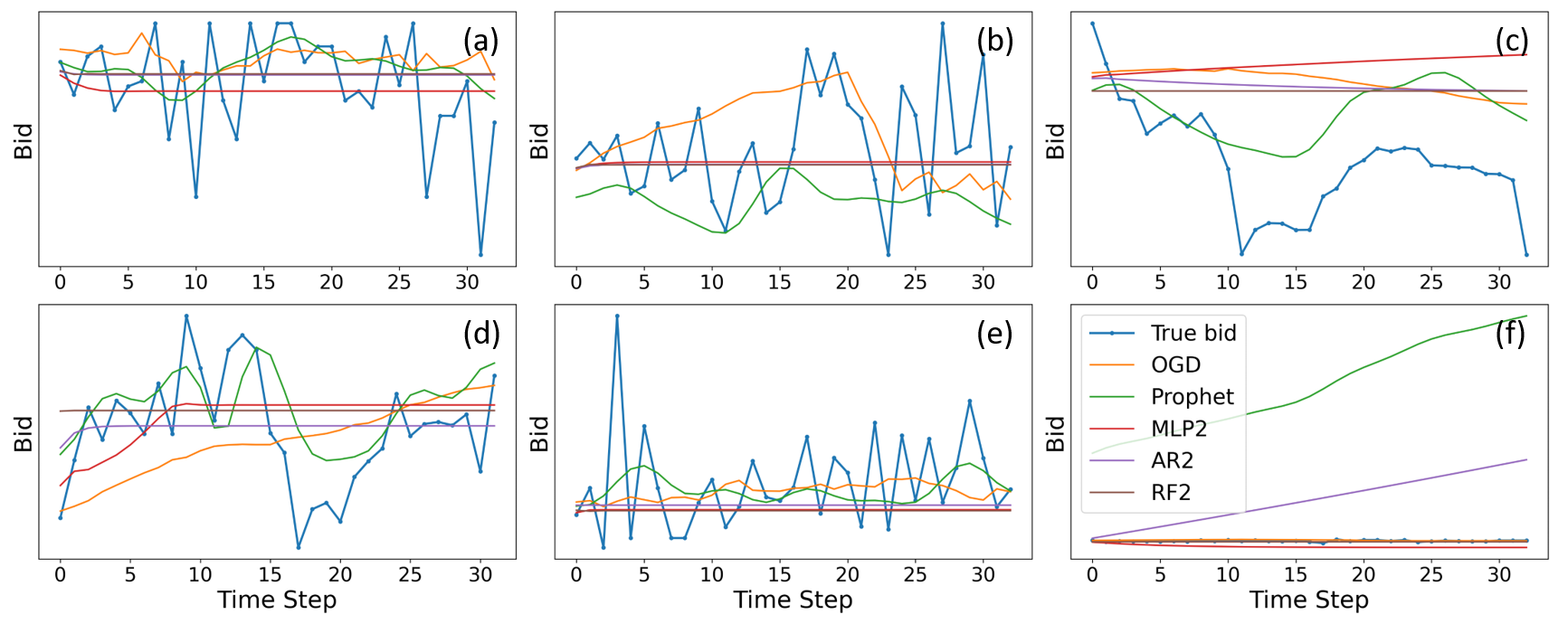}
\caption{Series Predictions}  
\label{fig:insample_curves}
\end{subfigure}
\centering
\begin{subfigure}{1.00\linewidth}
\includegraphics[width=1.00\linewidth]{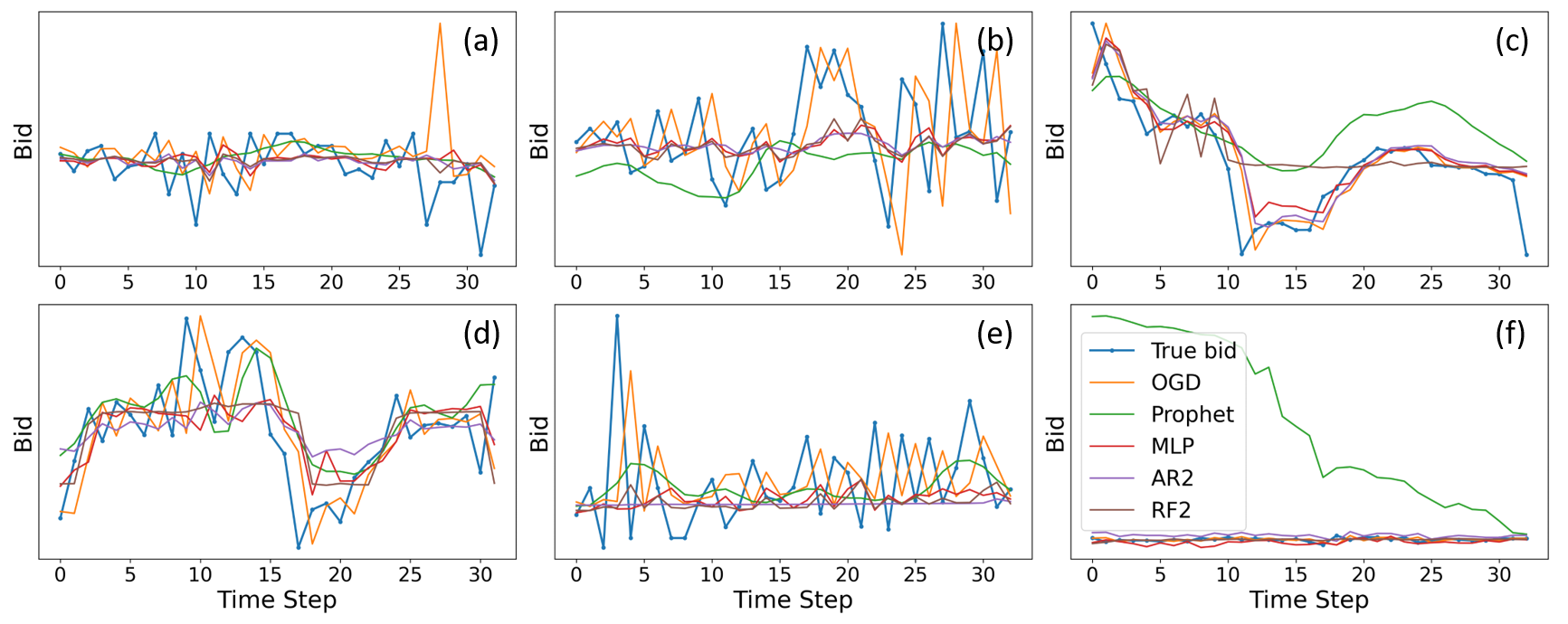}
\caption{Stepahead Predictions}  
\label{fig:insample_samples_stepahead}
\end{subfigure}
\caption{Example curves in the in-sample prediction setting (y-axis removed).}
\end{figure*}

Figures \ref{fig:insample_curves} and \ref{fig:insample_samples_stepahead}  show examples of bid curves and predictions of the OGD and the ML methods in the in-sample prediction setting.
The series prediction is a hard task %Projecting several steps into the future is a hard task 
since errors of a model are served as inputs to the next prediction steps and the errors may be accumulated. 
We see that Prophet manages to capture the bid dynamics well in most cases. E.g., Figure \ref{fig:insample_curves}(d) shows an impressive projection of Prophet 30 hours to the future in a non-trivial behavior. Also OGD usually manages to capture the correct direction of bid change. %In this 7d example, it catches the increasing bid trend except for the drop in bids that is expressed only in a flattening of the increase in predicted bids. 
In \ref{fig:insample_curves}(a) it is interesting to see that both Prophet and OGD outline similar curves, not far from the true bids, although the models are very different and are relying on very different inputs.
All models predict relatively smooth curves compared to the large hourly fluctuations observed in the bidding data.
The predictions of the non-linear RF2 and MLP2 and the linear AR2 are qualitatively similar, all usually ``cut'' the bid curves somewhere close to their average. The better MAPE score of RF2 shows that it is usually closer to the average than the other two methods. 
Figure \ref{fig:insample_samples_stepahead} shows the stepahead prediction for the same sample of 6 bidders. As can be seen, in the stepahead prediction task where the models are re-trained in each step with the true recent bid, the predictions of all models remain closer to the actual bid curve than in the series prediction task. Also the Prophet model, which is less suited for receiving step-by-step input, seems to benefit from this input and its predictions only get closer to the true bids with time; e.g. compare Prophet's predictions for example f in stepahead (Figure \ref{fig:insample_samples_stepahead}) and in series (Figure \ref{fig:insample_curves}) prediction tasks.

\FloatBarrier

%\subsection{Prediction Examples in the Co-variate Shift Prediction Setting} \label{subsec:curves_generalization}
\subsection{The Co-variate Shift Prediction Setting} \label{subsec:curves_generalization}

\begin{figure*}[h]
\centering
\begin{subfigure}{1.00\linewidth}
\includegraphics[width=1.00\linewidth]{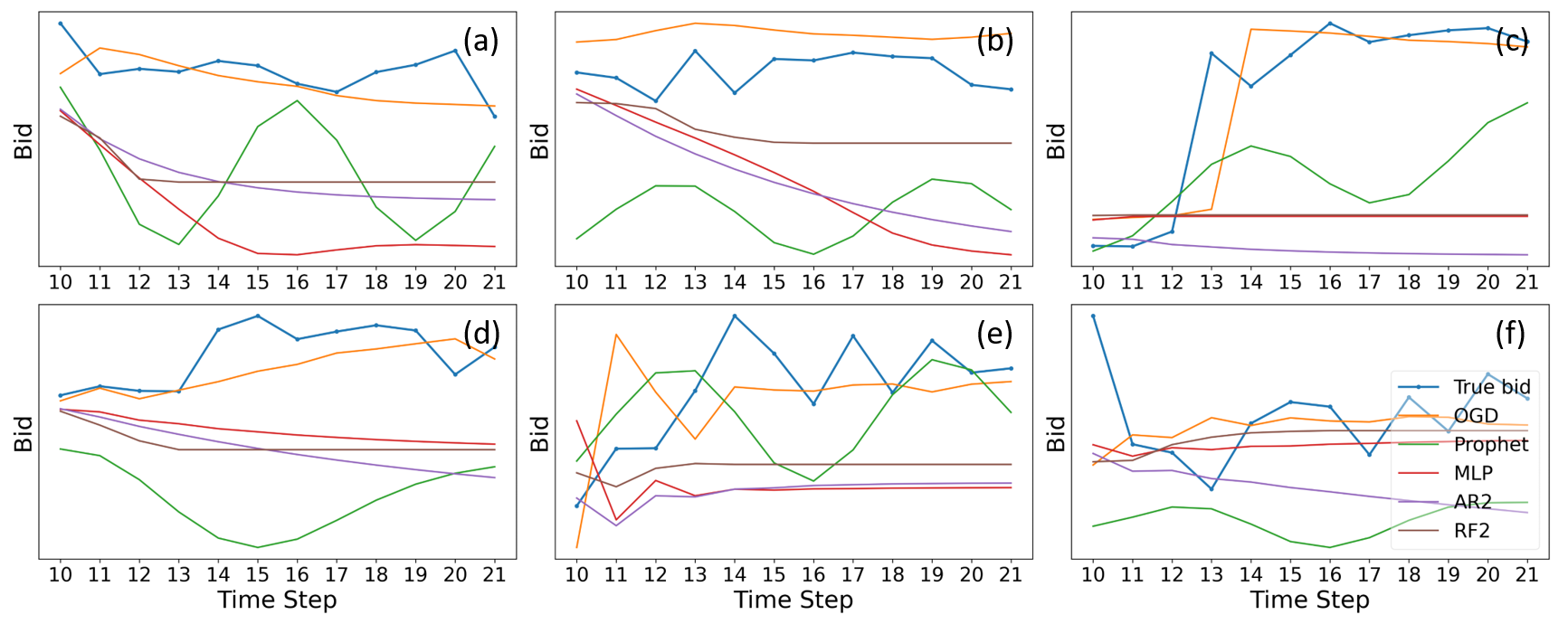}
\caption{Series Predictions}  
\label{fig:generalization_samples_series}
\end{subfigure}
\centering
\begin{subfigure}{1.00\linewidth}
\includegraphics[width=1.00\linewidth]{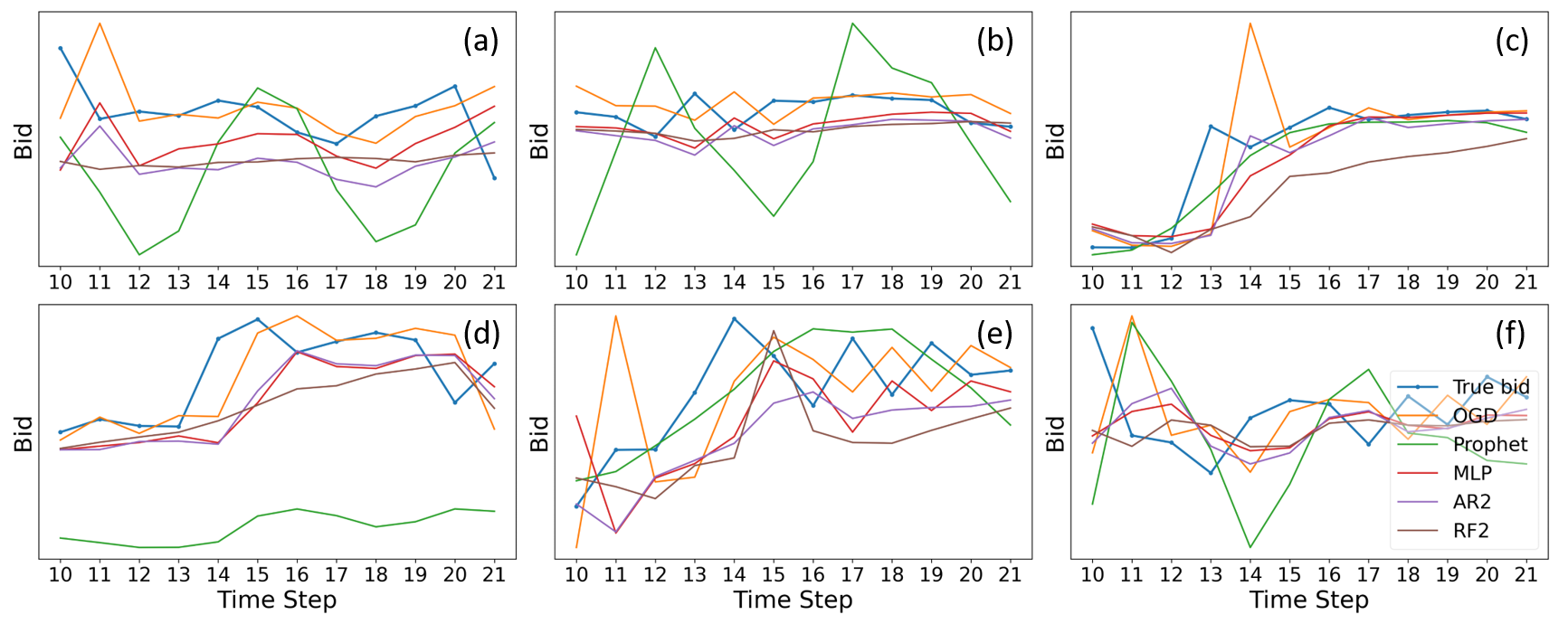}
\caption{Stepahead Predictions}  
\label{fig:generalization_samples_stepahead}
\end{subfigure}
\caption{Example curves in the co-variate shift prediction setting (y-axis removed).}
\end{figure*}

Figures \ref{fig:generalization_samples_series} and \ref{fig:generalization_samples_stepahead}  show examples of bid curves and predictions of the OGD and the ML methods in the co-variate shift prediction setting.
In the series prediction task (Figure \ref{fig:generalization_samples_series}), the plots demonstrate how OGD typically matches the new (higher) bid level of the test data even though it did not see bids from that bid distribution in the training data. In contrast, the ML methods fail to adapt to the new level of bids and their predictions typically remain close to the lower bids on which they were trained. The Prophet model tends to predict periodicity even when it is absent in the true data, possibly due to Fourier analysis on a relatively small training dataset for each player that may produce noise artifacts. For example, in \ref{fig:generalization_samples_series}(c), Prophet catches the increasing trend in bids, but still forces a periodic change.  RF2, MLP2, and AR2, that all receive the same input of the 2 previous bids, predict similar smooth dynamics, though RF2 and MLP2 seem somewhat closer to the higher bid levels of the test data than the linear AR2 model. 
Figure \ref{fig:generalization_samples_stepahead} shows prediction examples for the same %bidders 
sample of prediction days 
in the stepahead prediction task. In this more simple prediction task, where the models are re-trained at each step with the true recent bid, the predictions of all models are closer to the actual bid curve than in the series prediction task, though still the OGD predictions seem to match the new bid level better than the ML methods which usually predict lower than the true bid.

\begin{comment}
\textbf{Comments:} (a) The power-law-like distribution of $\alpha$ is likely an artifact of the power-law distribution of bids, utilities, etc. The distribution of $\alpha$ normalized by the value or the average utility of each player is likely to be more balanced and more informative about the distribution of the strength of the visibility bias. (b) It is likely that this shows that the $xmax$ is usually too high to represent actual clicks, and the parameter $vis_0$ fixes this to the right scale.
\end{comment}

\end{document}